\documentclass [12pt]{article}
\usepackage{amsmath,amssymb,cite}

\usepackage{url,hyperref}

\usepackage{tabularx}
\usepackage[english]{babel}
\usepackage[dvips]{graphicx}
\usepackage{indentfirst}
\usepackage{epsfig}

\begin{document}

\title{New Algorithm and Phase Diagram of Noncommutative $\Phi^4$ on the Fuzzy Sphere}

\author{ Badis Ydri\footnote{Email:ydri@stp.dias.ie,~badis.ydri@univ-annaba.org.} \\
 Institute of Physics, BM Annaba University,\\
BP 12, 23000, Annaba, Algeria.\\
}

\maketitle
\abstract{We propose a new algorithm for simulating noncommutative phi-four theory on the fuzzy sphere based on, i) coupling the scalar field to a $U(1)$ gauge field, in such a way that in the commutative limit $N\longrightarrow \infty$, the two modes decouple and we are left with pure scalar phi-four on the sphere, and ii) diagonalizing the scalar field by means of a $U(N)$ unitary matrix, and then integrating out the unitary group from the partition function.  The number of degrees of freedom in the scalar sector reduces, therefore, from $N^2$ to  the $N$ eigenvalues of the scalar field, whereas the dynamics of the $U(1)$ gauge field, is given by   $D=3$ Yang-Mills matrix model with a Myers term. As an application, the phase diagram, including the triple point, of noncommutative phi-four theory on the fuzzy sphere, is reconstructed  with  small values of $N$ up to $N=10$, and large numbers of statistics. 
}
\tableofcontents


\section{Introduction}
The goal of this article is to reconstruct by means of a (hopefully)  novel, and efficient Monte Carlo method the phase diagram of noncommutative phi-four on the fuzzy sphere. This was originally done  in \cite{GarciaFlores:2009hf}. The basic theory is given by the following two-parameter matrix model   
\begin{eqnarray}
S_0&=&{\rm Tr}_H\big(- a[L_a,{\Phi}]^2+b{\Phi}^2+c{\Phi}^4\big).\label{fundamental}
\end{eqnarray}
In this equation $L_a$ are the $SU(2)$ generators in the irreducible representation with spin $s=(N-1)/2$, ${\rm Tr}_H{\bf 1}=N$, $b$ is the mass parameter, and $c$ is the coupling constant. The parameter $a$ can always be chosen to be equal to $1$. There are three known phases in this model. The usual Ising transition between disorder and uniform order. A matrix transition between disorder and a non-uniform ordered phase, and a (very hard to observe) transition between uniform order and non-uniform order. The three phases meet at a triple point \cite{GarciaFlores:2009hf,GarciaFlores:2005xc}.  The non-uniform phase, in which { rotational invariance is spontaneously broken}, is simply absent in the commutative theory. The non-uniform phase is the analogue of the stripe phase observed on the Moyal-Weyl spaces \cite{Gubser:2000cd}, whereas the disorder-to-non-uniform-order transition is the generalization of the one-cut-to-two-cut transition, observed in  the Hermitian quartic matrix model \cite{Brezin:1977sv,Shimamune:1981qf}, to the fuzzy sphere. 

This is a highly non-trivial problem, which is due mainly, to the more complicated phase structure of matrix scalar phi-four. It involves transitions between vacuum states,  with very low probability distributions, and as a consequence, they are extremely difficult to  sample correctly with the Metropolis algorithm. In particular the non-uniform-to-uniform transition is virtually unobservable in ordinary Metropolis, due to the absence of tunneling between the identity matrix, corresponding to the uniform phase, and the other idempotent matrices, corresponding to the non-uniform phase. This means simply that the Metropolis updating procedure does not sample correctly, and equally, i.e. according to the Boltzmann weight, the entire phase space which includes an infinite number of vacuum states.  This was circumvented, in \cite{GarciaFlores:2009hf,GarciaFlores:2005xc}, by a complicated variant of the Metropolis algorithm, in which detailed balance is broken. This problem was also studied in \cite{Martin:2004un,Panero:2006bx,Das:2007gm,Medina:2007nv}. The analytic derivation of the phase diagram of noncommutative phi-four on the fuzzy sphere was attempted in \cite{O'Connor:2007ea,Saemann:2010bw,Polychronakos:2013nca}.

 The related problem of Monte Carlo simulation of noncommutative phi-four on the fuzzy torus, and the fuzzy disc was considered in \cite{Ambjorn:2002nj,Bietenholz:2004xs}, and \cite{Lizzi:2012xy} respectively.

The main strategy employed, in this article, towards a better resolution of this problem, is to reduce the model down to its eigenvalues, without actually altering it. This is achieved by: 
\begin{itemize}
\item[$1)$] coupling the scalar field to a $U(1)$ gauge field, in such a way, that in the commutative limit $N\longrightarrow \infty$, the two modes decouple completely, and thus we return to an ordinary phi-four theory, and 
\item[$2)$] diagonalizing the scalar field by means of a $U(N)$ gauge transformation, viz $\Phi=U\Lambda U^+$, and then integrating out the unitary matrices $U$ and $U^+$ from the path integral. 
\end{itemize}
In this algorithm, we thus trade off the Monte Carlo simulation of the unitary matrices $U$ and $U^+$, in the original model (\ref{fundamental}), with the Monte Carlo simulation of a $U(1)$ gauge field on the fuzzy sphere, which we know is very efficient using ordinary Metropolis \cite{Ydri:2012bq}.

The primary interest, of this article, is therefore Monte Carlo simulation of a noncommutative phi-four theory,  coupled to a $U(1)$ gauge field on the fuzzy sphere,  using the Metropolis algorithm with exact detailed balance. The scalar field transforms in the adjoint representation of the $U(1)$ gauge group, and as a consequence, the scalar and gauge degrees of freedom decouple in the commutative limit $N\longrightarrow \infty$. In other words, this theory becomes an ordinary phi-four theory  in the commutative limit. In this theory,  the usual scalar kinetic action $\sim -Tr[L_a,\Phi]^2$ is replaced with $\sim -Tr[X_a,\Phi]^2$, where $X_a$ is itself obtained by Monte Carlo simulation of an appropriate gauge action, which will be centered around $\sim L_a$, in the so-called fuzzy sphere phase\footnote{The behavior in the matrix phase is very different and is not treated in here.}. The pure gauge action is given by $D=3$ Yang-Mills action, with a Chern-Simons (Myers) term. For $b=c=0$ the full action is in fact  $D=4$ Yang-Mills action, with a Chern-Simons (Myers) term.

This article is organized as follows. In section $2$, we present the detail of the $U(1)$ gauge covariant noncommutative phi-four theory on the fuzzy sphere, and also explain the Metropolis algorithm employed in our Monte Carlo simulations. In section $3$, we report our first numerical results, on the phase diagram of noncommutative phi-four on the fuzzy sphere, using our new algorithm. We give independent measurements, of the three transition lines, discussed above, and then derive our estimation of the triple point. These results are obtained with small values of $N$ up to $N=10$, and large numbers of statistics. In section $4$, we give a construction of a one-parameter family of noncommutative phi-four models on the fuzzy sphere, which define, a regularization of duality covariant noncommutative phi-four on the Moyal-Weyl plane. We conclude in section $5$, with a brief summary, and outlook.

 \section{Model and Algorithm}
 \subsection{The Action}
Instead of the basic model (\ref{fundamental}), which is the primary interest in this article,  we consider a four matrix model given by the action
\begin{eqnarray}
S=S_g+S_m.
\end{eqnarray}
\begin{eqnarray}
S_g=N Tr\big(-\frac{1}{4}[X_a,X_b]^2+\frac{2i\alpha}{3}\epsilon_{abc} X_aX_bX_c\big)+NTr\big(M Tr (X_a^2)^2+\beta X_a^2\big).\label{Sg}
\end{eqnarray}
\begin{eqnarray}
S_m=-\frac{Na_0}{2}Tr[X_a,\Phi]^2+Tr V(\Phi).\label{Sm}
\end{eqnarray}

\begin{eqnarray}
V(\Phi)=r\Phi^2+u\Phi^4.
\end{eqnarray}
The fuzzy sphere phase is given by the background 
\begin{eqnarray}
X_a=\alpha\varphi L_a~,~\varphi=\frac{1+\sqrt{1+4\mu (1+m^2)}}{2(1+m^2)}.
\end{eqnarray}
The values $m^2=2c_2M$ and $\mu=-9\beta/\alpha^2$, of interest, are (with $c_2=(N^2-1)/4$ being the Casimir operator)
\begin{eqnarray}
1)~m^2=0~,~\mu=0~,~2)~m^2=c_2~,~\mu=\frac{2}{9}(2c_2-1).
\end{eqnarray}
In the remainder we will be interested in the first case. 

The first scaled parameter is \cite{Ydri:2012bq}
\begin{eqnarray}
\tilde{\alpha}=\alpha\sqrt{N}.
\end{eqnarray}
In the notation of \cite{GarciaFlores:2009hf}\footnote{We will also refer to this article as FDX.}, after replacing with $X_a=\alpha\varphi L_a$ in $S_m$,  we have  $a=\tilde{\alpha}^2\varphi^2a_0/2$, $b=r$ and $c=u$. The other scaled parameters are therefore given by 
\begin{eqnarray}
\tilde{b}=\frac{b}{aN^{\frac{3}{2}}}=\frac{2}{a_0\tilde{\alpha}^2\varphi^2N^{\frac{3}{2}}}r~,~\tilde{c}=\frac{c}{a^2N^2}=\frac{4}{a_0^2\tilde{\alpha}^4\varphi^4N^2}u.\label{scaling}
\end{eqnarray}
The dependence of the model on the coupling constant $a_0$ is fully taken into account by considering $\tilde{b}$ and $\tilde{c}$ instead of $b$ and $c$. The situation with the coupling constant $\tilde{\alpha}$ is more subtle. We expect that for large values of $\tilde{\alpha}$ the gauge sector $S_g$ describes a $U(1)$ gauge field on the fuzzy sphere, and as a consequence, the matter sector $S_m$ describes a (real) scalar field in the adjoint representation of the gauge group on the fuzzy sphere. More precisely we have in general $X_a=\alpha\varphi (L_a+A_a)$, where $A_a$ is the $U(1)$ gauge field which depends generically on $\tilde{\alpha}$. For large  values of $\tilde{\alpha}$, the gauge field is weakly coupled to the scalar field, and in the commutative limit $N\longrightarrow \infty$, the two fields become fully decoupled due to the commutator structure of the interaction. This is one of the main principles underlying our algorithm. Hence, the dependence of the model on the coupling constant $\tilde{\alpha}$ is also fully taken into account, in the limit $N\longrightarrow \infty$,  by considering $\tilde{b}$ and $\tilde{c}$ instead of $b$ and $c$. The theory $S_g+S_m$ describes therefore, for large values of  $\tilde{\alpha}$ and large values of $N$, a scalar phi-four on the fuzzy sphere. 

In all of the simulations reported in this article, we take $a_0=1$ and $\tilde{\alpha}=10$ for concreteness. The choice for $\tilde{\alpha}$ is dictated by the fact that a fuzzy sphere phase, in the model with $r=u=0$ (the four dimensional Yang-Mills action), is known to persist only for values of $\tilde{\alpha}$ given by \cite{Ydri:2012bq}\footnote{The scalar sector is strictly speaking independent of  the parameter $a_0$ for all $N$, whereas it is independent of $\tilde{\alpha}$ only in the limit $N\longrightarrow \infty$. We can use values of $\tilde{\alpha}$ near $\tilde{\alpha}_*$, which corresponds to  large $A_a$, in order to enhance the contribution of the kinetic scalar action, and hence, excite the system to tunnel to the true minimum in each phase. This idea is not investigated thoroughly here. }
\begin{eqnarray}
\tilde{\alpha}\geq \tilde{\alpha}_*=2.55\pm 0.1.
\end{eqnarray}
Let us discuss the phase structure of the pure potential model $V(\Phi)$. The ground state configurations  are given by the matrices
\begin{eqnarray}
{\Phi}_0=0.
\end{eqnarray}
\begin{eqnarray}
{\Phi}_{\gamma}=\sqrt{-\frac{r}{2u}}U\gamma U^+~,~{\gamma}^2={\bf 1}_N~,~UU^+=U^+U={\bf 1}_N.
\end{eqnarray}
We compute $V[{\Phi}_0]=0$ and $V[{\Phi}_{\gamma}]=-r^2/4u$. The first configuration corresponds to the disordered phase characterized by $<{\Phi}>=0$. The second solution makes sense only for $r<0$, and it corresponds to the ordered phase characterized by $<{\Phi}>=\sqrt{-\frac{r}{2u}}U\gamma U^+$. There is a nonperturbative transition between the two phases which occurs, not at $r=0$, but at $r=r_*=-2\sqrt{Nu}$, which is known as the one-cut-to-two-cut transition\footnote{In terms of $\tilde{b}$ and $\tilde{c}$ the critical value occurs at $\tilde{b}=-2\sqrt{\tilde{c}}$. If the relation between $r_*$ and $u$ were on the other hand linear, viz $r_*\sim u$, then we would have instead $\tilde{b}/\sqrt{N}\sim \tilde{c}$. } . The idempotent $\gamma$ can always be chosen such that $\gamma=\gamma_k={\rm diag}({\bf 1}_{k},-{\bf 1}_{N-k})$. The orbit of $\gamma_k$ is the Grassmannian manifold $U(N)/(U(k)\times U(N-k))$, the dimension of which is $d_k=2kN-2k^2$. It is not difficult to show that this dimension is maximum at $k=N/2$ (assuming that $N$ is even), and hence from entropy argument, the most important two-cut solution is the so-called stripe configuration given by $\gamma={\rm diag}({\bf 1}_{{N}/{2}},-{\bf 1}_{{N}/{2}})$.

In the theory given by the action $S_m$, we have therefore three possible phases. The phase  characterized by the expectation value $<{\Phi}>=0$, the phase characterized by  $<{\Phi}>=\pm \sqrt{-r/2u}~{\bf 1}_N$,  and the phase characterized by $<{\Phi}>=\pm\sqrt{-r/2u}~\gamma$, where $ \gamma=({\bf 1}_{N/2},-{\bf 1}_{/2})$. We use the terminology
\begin{eqnarray}
&&<{\Phi}>=0~~{\rm disordered}~{\rm phase}.
\end{eqnarray}
\begin{eqnarray}
&&<{\Phi}>=\pm\sqrt{-\frac{r}{2u}}{\bf 1}_N~~{\rm Ising}~({\rm uniform})~{\rm phase}.\label{ising}
\end{eqnarray}
\begin{eqnarray} 
&&<{\Phi}>=\pm\sqrt{-\frac{r}{2u}}\gamma~~{\rm matrix}~({\rm nonuniform}~{\rm or}~{\rm stripe})~{\rm phase}.\label{matrix}
\end{eqnarray}
There are therefore three possible phase transitions, and as a consequence, there exists a triple point. The famous $2$nd order Ising phase transition $0\longrightarrow \pm\sqrt{-r/2u}~{\bf 1}_N$. The famous $3$rd order matrix phase transition $0\longrightarrow \pm\sqrt{-r/2u}({\bf 1}_{N/2},-{\bf 1}_{N/2})$. Clearly then, there must exist also a transition between the Ising and matrix configurations, viz ${\bf 1}\longrightarrow \gamma$, which is expected to be a continuation of the Ising line to large values of the coupling constant $u$, and thus it is expected to be $2$nd order.

In the numerical simulations, we will be interested in the values $m^2=\mu=0$. As a test of our simulations, we will use the following exact Schwinger-Dyson identity\footnote{By changing $X_a$ to $X_a^{'}=(1+\epsilon)X_a$ and ${\Phi}$ to ${\Phi}^{'}=(1+\epsilon){\Phi}$, in the partition function, we can derive from the invariance of the path integral this identity.}
    \begin{eqnarray}
<{\rm IDE}>&=&4N^2.\label{ide}
\end{eqnarray}
The operator ${\rm IDE}$ is given by
      \begin{eqnarray}
{\rm IDE}&=&4N Tr\big(-\frac{1}{4}[X_a,X_b]^2\big)+3N Tr\big(\frac{2i\alpha}{3}\epsilon_{abc} X_aX_bX_c\big)+4\big(-\frac{N}{2}Tr[X_a,\Phi]^2\big)\nonumber\\
&+&2rTr\Phi^2+4uTr\Phi^4.
\end{eqnarray}

\subsection{Algorithm and Simulation}
The path integral we want to simulate is 
   \begin{eqnarray}
Z&=&\int \prod_a dX_a\int d\Phi~\exp\bigg[-N Tr\big(-\frac{1}{4}[X_a,X_b]^2+\frac{2i\alpha}{3}\epsilon_{abc} X_aX_bX_c\big)+\frac{N}{2}Tr[X_a,\Phi]^2\nonumber\\
&-&Tr \big(r\Phi^2+u\Phi^4\big)\bigg].
\end{eqnarray}
Let us now diagonalize the hermitian $N\times N$ matrix $\Phi$ by writing the polar decomposition  $\Phi=U^+\Lambda U$, $\Lambda={\rm diag}(\lambda_1,....,\lambda_N)$ for unitary $N\times N$ matrices $U$. The measure becomes
\begin{eqnarray}
d\Phi = [dU]  \prod_{i=1}^N d\lambda_i \Delta_N(\lambda)~,~ \Delta_N(\Lambda)= \prod_{1\leq i<j \leq N} (\lambda_i-\lambda_j)^2. 
\end{eqnarray}
In above $[dU]$ is  the Haar measure on the
group $U(N)$, whereas ${\Delta}_N(x)$ is the  Vandermonde determinant. By using now gauge invariance of the above path integral, we can reabsorb  the unitary matrix $U$, by changing $X_a$ as $ X_a\longrightarrow UX_aU^+$, and as a consequence, the integral over $U$ decouples. The path integral becomes then  
  \begin{eqnarray}
Z&=&\int \prod_a dX_a\int d\Lambda~\exp\bigg[-N Tr\big(-\frac{1}{4}[X_a,X_b]^2+\frac{2i\alpha}{3}\epsilon_{abc} X_aX_bX_c\big)+\frac{N}{2}Tr[X_a,\Lambda]^2\nonumber\\
&-&Tr \big(r\Lambda^2+u\Lambda^4\big)+\ln \Delta_N(\Lambda)\bigg].
\end{eqnarray}
The scalar action is, then, given by
 \begin{eqnarray}
S[\Lambda]&=&-\frac{N}{2}Tr[X_a,\Lambda]^2+Tr \big(r\Lambda^2+u\Lambda^4\big)-\ln \Delta_N(\Lambda)\nonumber\\
&=&-N\sum_{ij}(X_a)_{ij}(X_a)_{ji}\lambda_i\lambda_j+N\sum_i(X_a^2)_{ii}\lambda_i^2+\sum_i(r\lambda_i^2+u\lambda_i^4)-\sum_{i\neq j}\ln|\lambda_i-\lambda_j|.\nonumber\\\label{Smdi}
\end{eqnarray}
We will apply the Metropolis algorithm in which we change the eigenvalues $\lambda_i$ one at a time. Under the change of the eigenvalue $\lambda_i$ (fixed $i$), i.e. under $\lambda_n\longrightarrow \lambda_n^{'}=\lambda_n+\delta_{ni}\epsilon$, the action $S[\Lambda]$ changes as
 \begin{eqnarray}
\Delta S_i[\Lambda]&=&2N\epsilon(X_a^2)_{ii}\lambda_i+N(X_a^2)_{ii}\epsilon^2-2N\epsilon\sum_n(X_a)_{ni}(X_a)_{in}\lambda_n-N(X_a)_{ii}^2\epsilon^2\nonumber\\
&+&r(\epsilon^2+2\epsilon\lambda_i)+u(\epsilon^2+2\epsilon\lambda_i)(\epsilon^2+2\epsilon\lambda_i+2\lambda_i^2)-2\sum_{j\neq i}\ln |1+\frac{\epsilon}{\lambda_i-\lambda_j}|.
\end{eqnarray}
The first line is the variation of the kinetic term, the two first terms of the second line provide the variation of the potential, whereas the last term is the variation of the  Vandermonde determinant.

The variation of the action $S[\Lambda]$, under the change of the entry $(i,j)$ of  one of the matrices $X_a$, say $X_a\longrightarrow X_a+\Delta X_a$, is given by

 \begin{eqnarray}
\Delta S_{a,(i,j)}[\Lambda]&=&2N\sum_{n,m}(X_a)_{nm}(\Delta X_a)_{mn}(\lambda_n^2-\lambda_n\lambda_m)\nonumber\\
&+&N\sum_{n,m}(\Delta X_a)_{nm}(\Delta X_a)_{mn}(\lambda_n^2-\lambda_n\lambda_m).
\end{eqnarray}
We choose
\begin{eqnarray}
(\Delta X_a)_{mn}=\delta_{ni}\delta_{mj}\epsilon^*+\delta_{nj}\delta_{mi}\epsilon.
\end{eqnarray}
The variation becomes
 \begin{eqnarray}
\Delta S_{a,(i,j)}[\Lambda]&=&2N(X_a)_{ij}\epsilon^*(\lambda_i^2-\lambda_i\lambda_j)+2N(X_a)_{ji}\epsilon(\lambda_j^2-\lambda_i\lambda_j)+2N\epsilon\epsilon^*(\lambda_i-\lambda_j)^2.\nonumber\\
\end{eqnarray}
We remark that for diagonal elements, i.e. $i=j$, this variation vanishes identically. This is simply due to the fact that the scalar kinetic action does not depend on diagonal elements of the matrices $X_a$. The full variation   under the change of the entry $(i,j)$ of  one of the matrices $X_a$, which will enter the Metropolis algorithm, will naturally contain contributions coming from the pure gauge action. This part has been used elsewhere with great success \cite{Ydri:2012bq}.

The identity in this case still reads as in (\ref{ide}), with the operator ${\rm IDE}$  given by
      \begin{eqnarray}
{\rm IDE}&=&4N Tr\big(-\frac{1}{4}[X_a,X_b]^2\big)+3N Tr\big(\frac{2i\alpha}{3}\epsilon_{abc} X_aX_bX_c\big)+4\big(-\frac{N}{2}Tr[X_a,\Lambda]^2\big)\nonumber\\
&+&2rTr\Lambda^2+4uTr\Lambda^4.
\end{eqnarray}
The Vandermonde action contributes to the integer $4N^2$, and as a consequence, it does not appear in ${\rm IDE}$.

It is very hard  to generate, in the simulation, a sample of gauge and scalar configurations which satisfy this exact identity, due to the large degree of auto-correlation observed in the fuzzy sphere phase, i.e. for large values of $\tilde{\alpha}$. To reduce this undesirable effect, we separate any two successive configurations used in our measurements,  by a large number of unused Monte Carlo configurations.

We measure the expectation value of the action $<S_m>$, the total power $P_T$, the power in the zero mode $P_0$, the kinetic term $<K>$, the specific 
  heat $C_v$ \footnote{In the formula of the specific heat the action does not include the Vandermonde.}, the magnetization $M$ and the susceptibility $\chi$. The action has already been defined . The other observables are defined by

\begin{eqnarray}
K=-\frac{N}{2}<Tr[X_a,\Lambda]^2>.
\end{eqnarray}

\begin{eqnarray}
C_v=<S^2>-<S>^2.
\end{eqnarray}
\begin{eqnarray}
M=<|Tr\Lambda|>.
\end{eqnarray}
\begin{eqnarray}
\chi=<|Tr {\Lambda}|^2>-<|Tr{\Lambda}|>^2.
\end{eqnarray}
\begin{eqnarray}
P_0=<(Tr{\Lambda})^2>/N^2.
\end{eqnarray}
\begin{eqnarray}
P_T=<Tr{\Lambda}^2>/N.
\end{eqnarray}
We use the Metropolis algorithm to update configurations, and we use the jackknife method to estimate error bars. The choice of the initial state is irrelevant. The Metropolis algorithm and the initial state used are discussed below in more detail. Typically, starting from a given/prepared initial state we run the Metropolis algorithm for $T_{\rm T}$ thermalization steps to achieve thermalization, and $T_{\rm MC}$ Monte Carlo steps for the actual Monte Carlo evolution. We record all of the $T_{\rm MC}$ configurations and compute averages over them. Each two successive Monte Carlo steps are separated by $T_{\rm C}$ auto-correlation steps. The value of $T_{\rm C}$ can be chosen to be at least equal to the auto-correlation time, for a given set of parameters, which can be computed using the usual formula.

\section{The Phase Diagram}

\subsection{The Ising Phase Transition}  

In this case, the Metropolis updating procedure consists  in going through the entries of each matrix $X_a$,  and through each of the eigenvalues of $\Lambda$,  sequentially, and then attempting to change them in the usual way. 

The initial state is prepared as follows. First, we start from $\Lambda=0$ and $X_a=\alpha L_a$, at $\tilde{b}=0$,  which we know is the true minimum at this point, and then run a Metropolis updating procedure, on this initial state keeping $X_a$ fixed, without taking into account the effect of the Vandermonde determinant, which is obviously the hardest part to thermalize,   to obtain the actual initial state for $\tilde{b}=0$. Using this initial state, we launch the full Metropolis updating procedure. 

Next, we start changing $\tilde{b}$ adiabatically (slowly), in such a way that the initial configuration for each new value of $\tilde{b}$ is the last configuration obtained for the previous  value of $\tilde{b}$. Each time, we run starting from this initial state, a Metropolis updating procedure, keeping $X_a$ fixed, and without the effect of the Vandermonde determinant, to obtain the actual initial state for that particular value of $\tilde{b}$, before we launch the full Metropolis updating procedure. 

We have checked that the location of the  disordered-to-uniform-ordered transition does not depend on the above procedure, and thus it is fully independent of the initial conditions utilized.

A simulation consists typically of $2T_{\rm T}+T_{\rm C}\times T_{\rm MC}$ steps where $T_{\rm MC}=T_{\rm T}=2^{13} (N=4,6)$ or $T_{\rm MC}=T_{\rm T}=2^{14} (N=10)$, and $T_{\rm C}=2^5$. The first $T_{\rm T}$ steps is done at fixed $X_a=\alpha L_a$, and without the Vandermonde determinant.

We have verified that the identity (\ref{ide}) holds within statistical errors. More precisely, we have only  admitted data points satisfying $<{\rm IDE}>/N^2=4.00\pm 0.25$ ($N=6,10$) and $<{\rm IDE}>/N^2=4.00\pm 0.30$ ($N=4$).

\begin{figure}[htbp]
\begin{center}
\includegraphics[width=5.9cm,angle=-90]{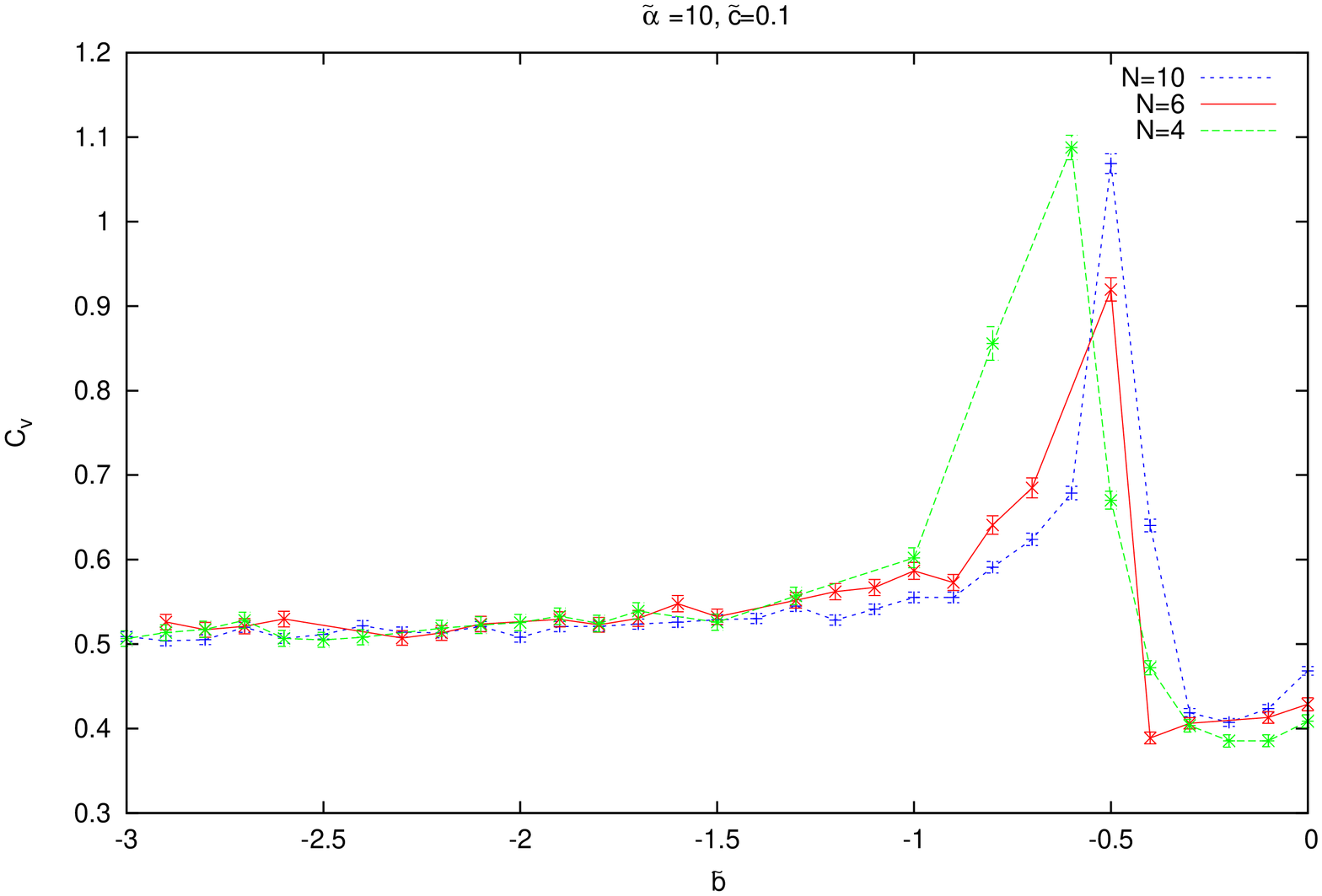}
\includegraphics[width=5.9cm,angle=-90]{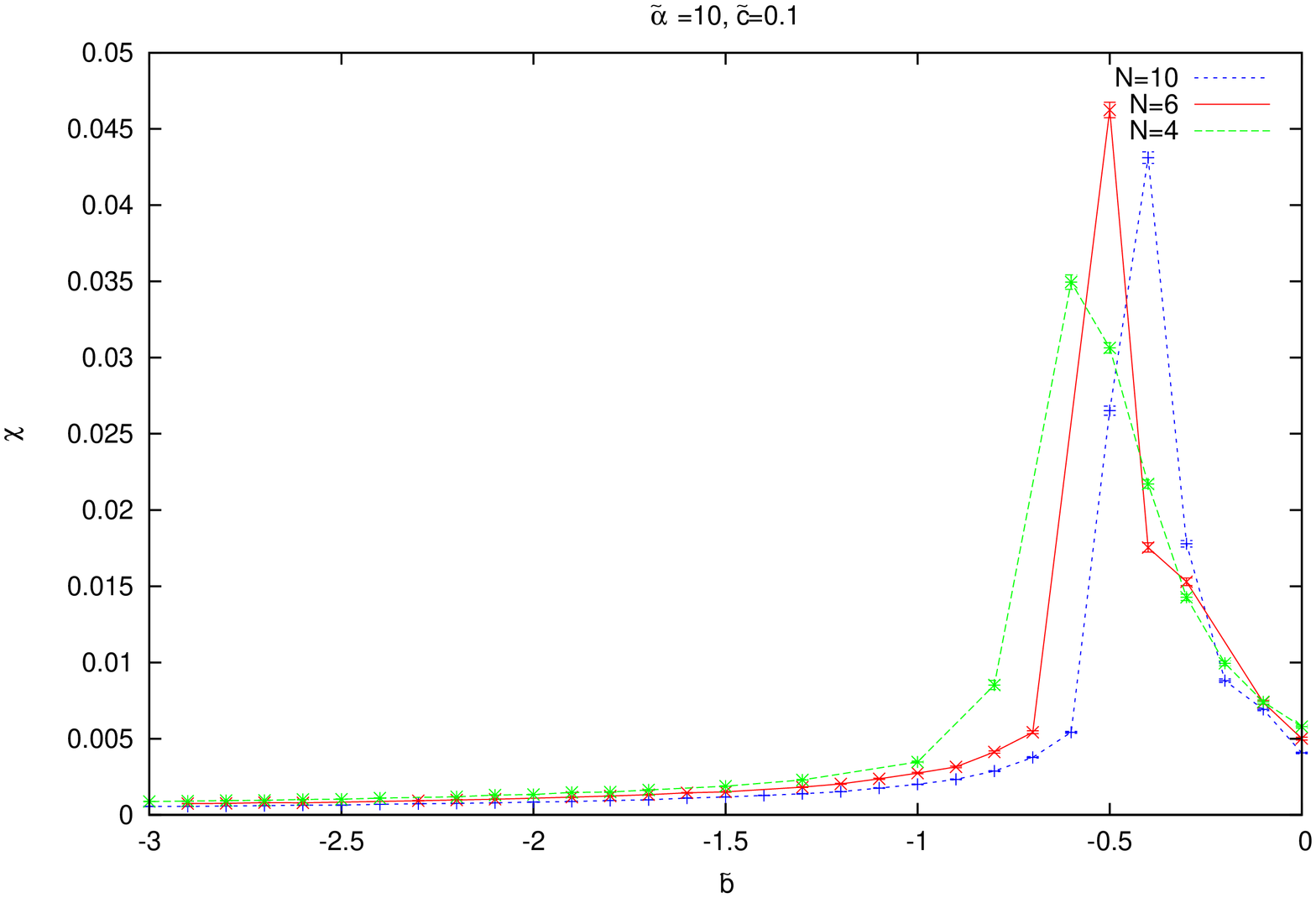}
\includegraphics[width=5.9cm,angle=-90]{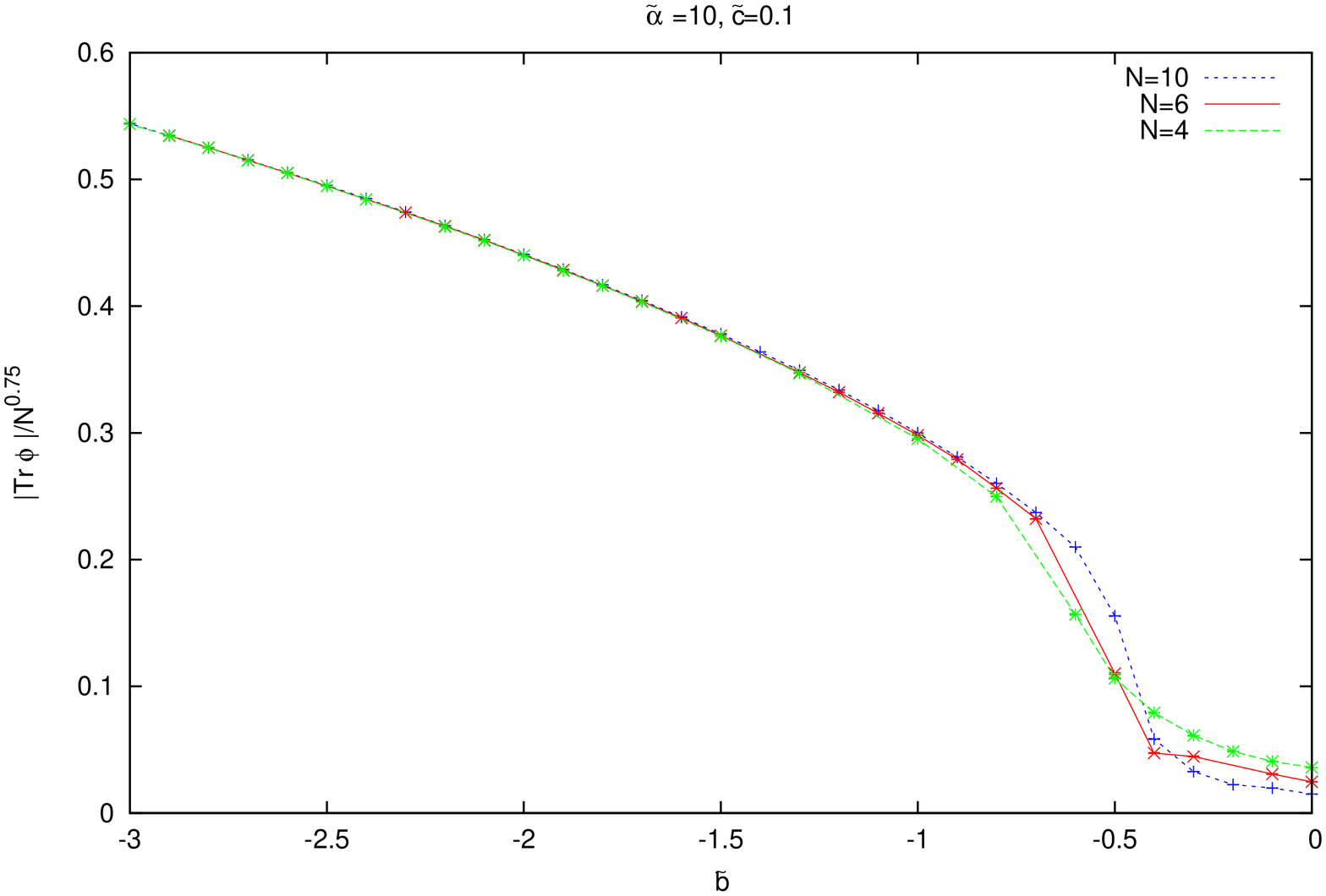}
\caption{ The disordered-to-uniform-ordered phase transition. }\label{phase1}
\end{center}
\end{figure}

The disordered-to-uniform-ordered transition is shown on figure (\ref{phase1}).
This transition can appear only for small values of $\tilde{c}$. We take for example $\tilde{c}=0.1$. 
The $2$nd order Ising transition (location of the peaks in the specific heat and the susceptibility) occurs at $\tilde{b}_*=-0.5\pm 0.1$ (for $C_v$, $N=10$),  $\tilde{b}_*=-0.4\pm 0.1$ (for $\chi$, $N=10$), $\tilde{b}_*=-0.63\pm 0.13$ (for $C_v,\chi$, $N=6$) and  $\tilde{b}_*=-0.53\pm 0.13$ (for $C_v,\chi$, $N=4$).  If we take the arithmetic average of the values obtained from the specific heat and the susceptibility for different $N$, as an estimation of the location of the Ising transition, we obtain for   $\tilde{c}=0.1$ the value $-0.54\pm 0.12$. These results  for  $\tilde{c}=0.1$, and those for  $\tilde{c}=0.3$, are included in table (\ref{tableI}).
 
 \begin{table}[h]
\centering
\begin{tabular}{|l|c|c|c|c| }
\hline
$\tilde{c}$ & $N=10$ & $N=6$ &  $N=4$ & $\tilde{b}_* ({\rm arithmetic}~{\rm average})$ \\
\hline 
$0.1$ &  $-0.45\pm 0.1$ & $-0.53\pm 0.13 $ & $-0.63\pm 0.13$ &  $-0.54\pm 0.12$\\ 
$0.3$ &  $-1.5\pm 0.2$ & $-1.53\pm 0.33 $ & $-1.53\pm 0.33$ &  $-1.52\pm 0.29$\\ 
\hline 
\end{tabular}
\caption{The  Ising transition points. }\label{tableI}
\end{table}

Using just these two points, we can determined the boundary between the  disordered and the uniform-ordered phases, as a straight line, with slope given by
 \begin{eqnarray}
{\rm slope}=\frac{0.3-0.1}{-1.54-(-0.54)}=-0.2.
\end{eqnarray}
The fit to the  uniform-ordered-to-disordered transition line is given by (suppressing error bars because they are quite insignificant in this case) 
 \begin{eqnarray}
\tilde{c}=-0.2\tilde{b}.\label{ising}
\end{eqnarray}
This agrees with  \cite{GarciaFlores:2009hf}. We note that we have dropped out the intercept in the fit equation because it is, within statistical errors,  completely negligible.  This confirms the general expectation that the Ising line must go through the origin $(\tilde{c},\tilde{b})=(0,0)$.

We note finally that this transition can also be obtained using the usual Metropolis algorithm with the ordinary pure scalar action, i.e. with the action (\ref{Sm}), with $X_a$ fixed given by $X_a=\alpha L_a$.

\subsection{The Uniform-to-Non-Uniform Phase Transition}  

\paragraph{Thermalization and Tunneling:}
The non-uniform-ordered-to-uniform-ordered transition can appear only for medium and large values of $\tilde{c}$. It is a second order phase transition, which is the continuation of the Ising transition, to larger values of $\tilde{c}$.



The non-uniform-ordered phase is the phase associated with spontaneous breaking of rotational/translational symmetry on the fuzzy sphere\footnote{Under a unitray transformation $U$ the idempotent $\gamma$ transforms as $\gamma \longrightarrow U\gamma U^+$. For $\gamma=0$ (disorder) and $\gamma=\pm {\bf 1}_N$ (uniform) we obtain rotational/translational invariance.}. This fact lies at the heart of its fundamental importance.

Firstly, we note that this transition is virtually impossible to be observed using the Metropolis algorithm, with the action (\ref{Sm}), where $X_a=\alpha L_a$.

We can probe the  uniform-to-non-uniform phase transition (although still very difficult), using the Metropolis algorithm with the action (\ref{Smdi}), where $X_a$ is obtained itself via the Metropolis algorithm, with the action (\ref{Sg}), where $M=\beta=0$. 

The initial state, for a fixed $\tilde{b}$ and $\tilde{c}$, is prepared as follows. We start from a random configuration $\Lambda$, and from $X_a=\alpha L_a$, and then run a Metropolis updating procedure for $T_{\rm T}$ steps on this initial state, without taking into account the effect of the Vandermonde determinant at fixed $X_a$, to obtain the actual initial state.  Starting from this resulting state, we run a  full Metropolis updating procedure for  $T_{\rm T}+T_{\rm C}\times T_{\rm MC}$ steps. This whole process consists a single simulation.

A simulation consists typically of $2T_{\rm T}+T_{\rm C}\times T_{\rm MC}$ steps, where $T_{\rm MC}=T_{\rm T}=2^{13}$, and $T_{\rm C}=2^5$ for $N=6$, and $T_{\rm MC}=T_{\rm T}=2^{14}$ and $T_{\rm C}=2^6$ for $N=8$. 

Only simulations satisfying $<{\rm IDE}>/N^2=4.00\pm 0.25$ ($N=6,8$)  are admitted in accordance with the Schwinger-Dyson identity (\ref{ide}).

We have studied thermalization in great detail. Typically, we tend to repeat the same simulation  $T_{\rm S}=2^{7}+1$ times, where each simulation is started from the final state obtained in the previous simulation. The goal is to assess tunneling transitions between the different vacua $<\Phi>\sim {\bf 1}$, $\gamma$ and $\gamma_k$. 

As pointed out earlier the vacuum state $<\Phi>\sim{\bf 1}$ has always the smallest energy, whilst the vacuum state  $<\Phi>\sim \gamma$  has always the largest energy. The other states are naturally somewhere in between. However, from entropy considerations, it is the state  $<\Phi>\sim \gamma$, which has the largest phase space volume, which can be seen from the size of the Grassmannian manifold $U(N)/(U(k)\times U(N-k))$, given by the dimension $d_k=2kN-2k^2$, which is maximal for $k=N/2$. 

At infinite $N$, we therefore expect that only $<\Phi>\sim{\bf 1}$ and $<\Phi>\sim \gamma$ are stable vacua and thus must be observed, while for finite $N$, tunneling transitions to other states are expected and will in fact also be observed.

\begin{figure}[htbp]
\begin{center}
\includegraphics[width=5.9cm,angle=-90]{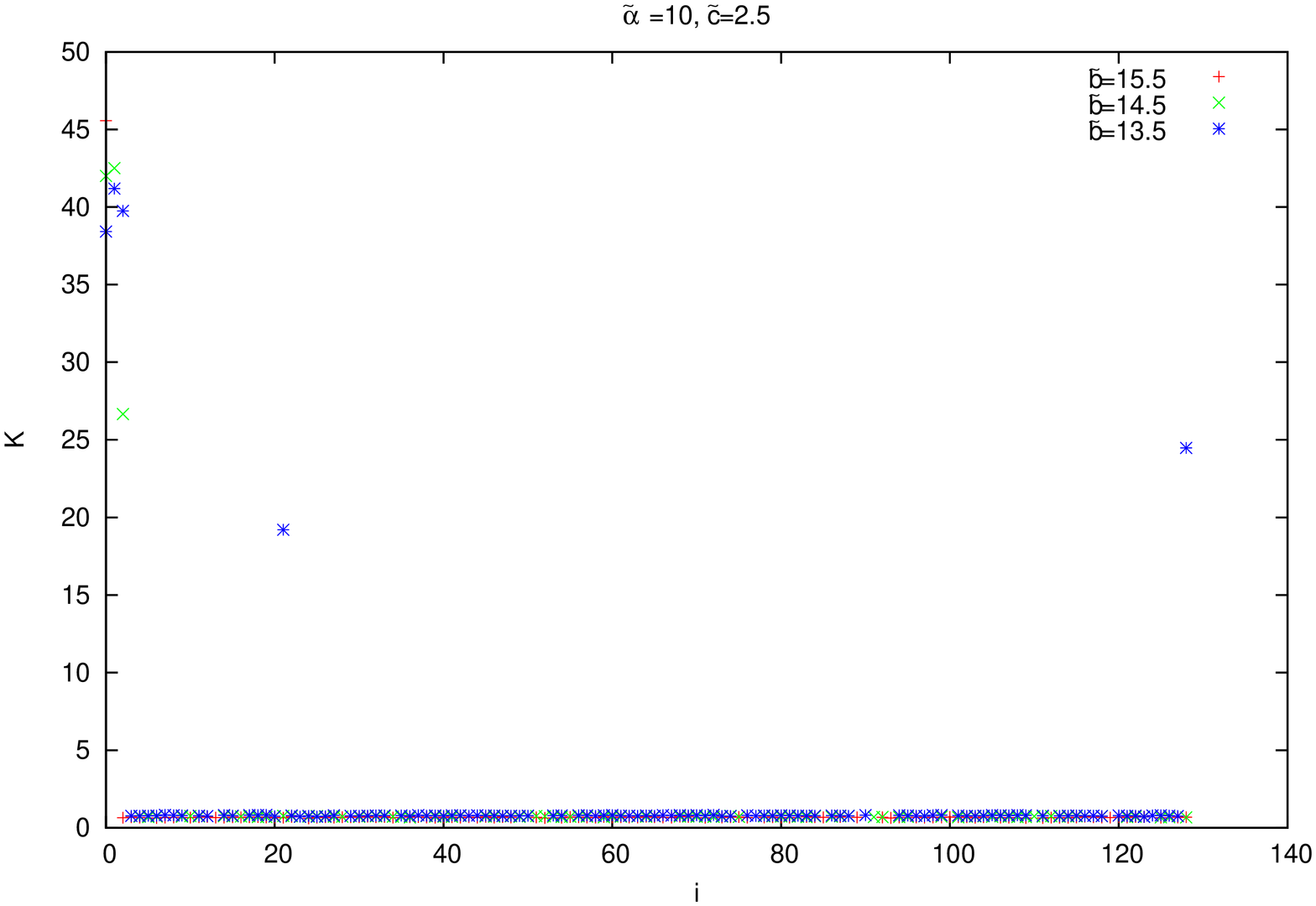}
\includegraphics[width=5.9cm,angle=-90]{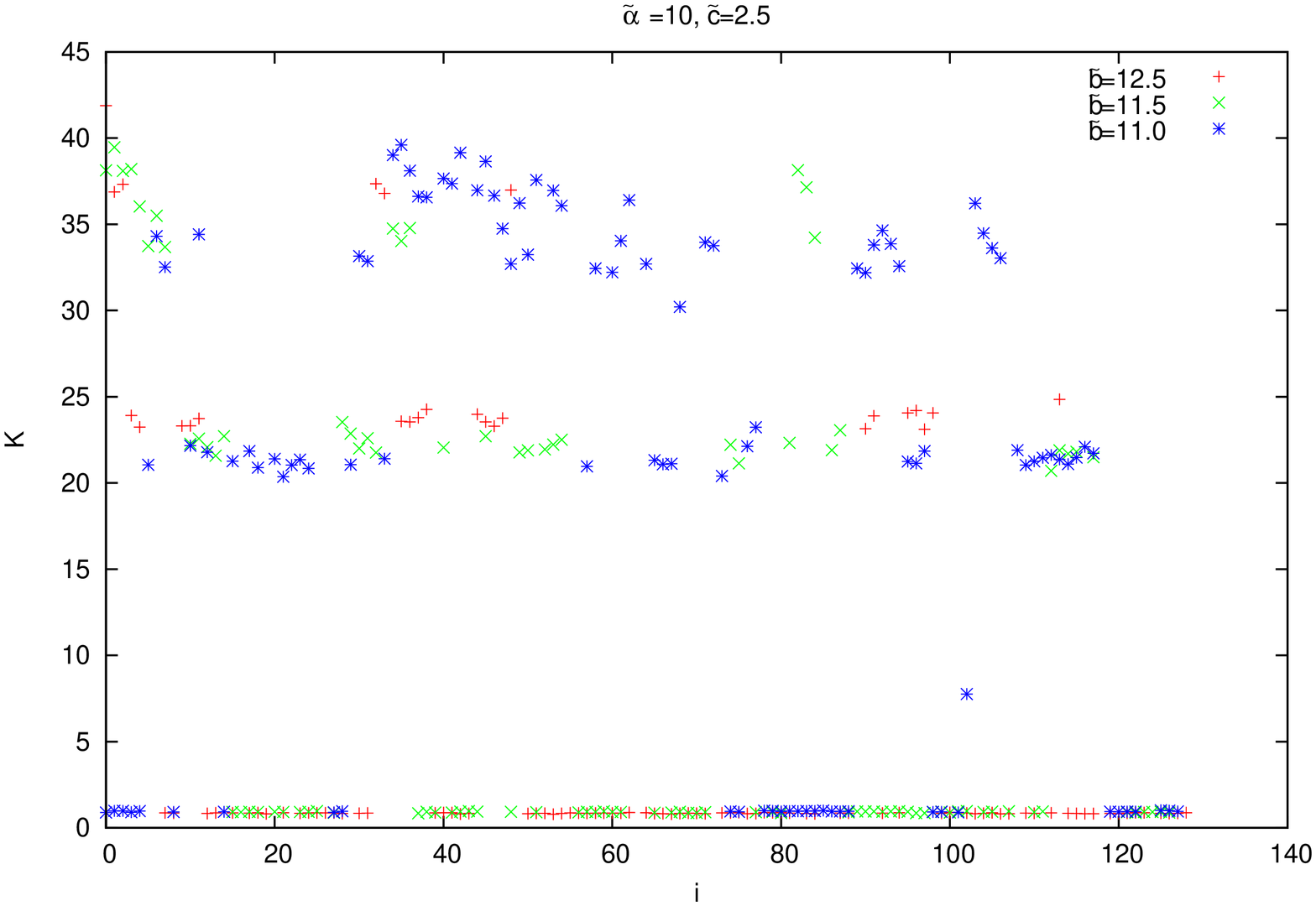}
\includegraphics[width=5.9cm,angle=-90]{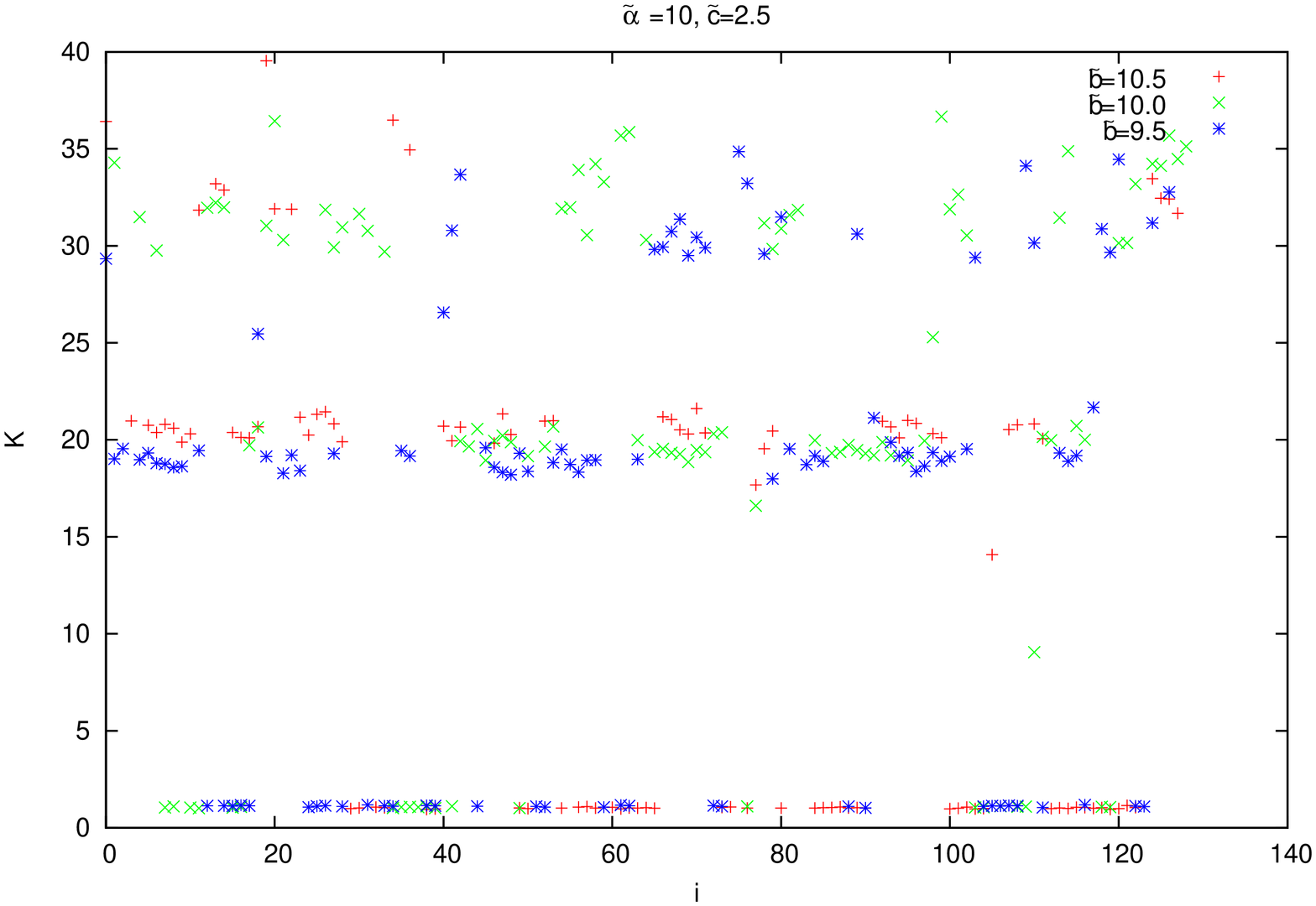}
\includegraphics[width=5.9cm,angle=-90]{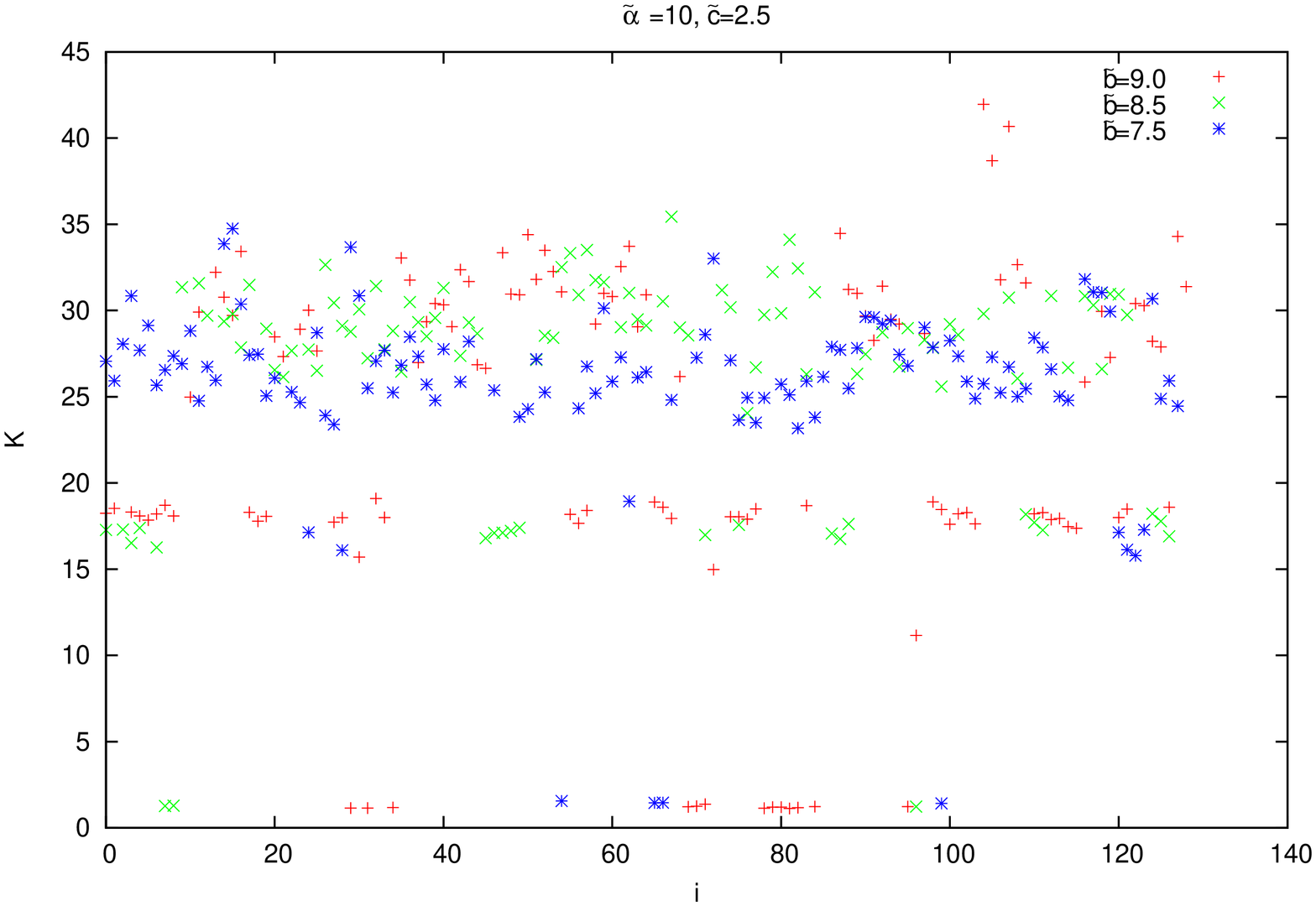}
\caption{Thermalization of the kinetic action across the non-uniform-ordered-to-uniform-ordered phase transition. }\label{phase2p0}
\end{center}
\end{figure}

\begin{figure}[htbp]
\begin{center}
\includegraphics[width=5.9cm,angle=-90]{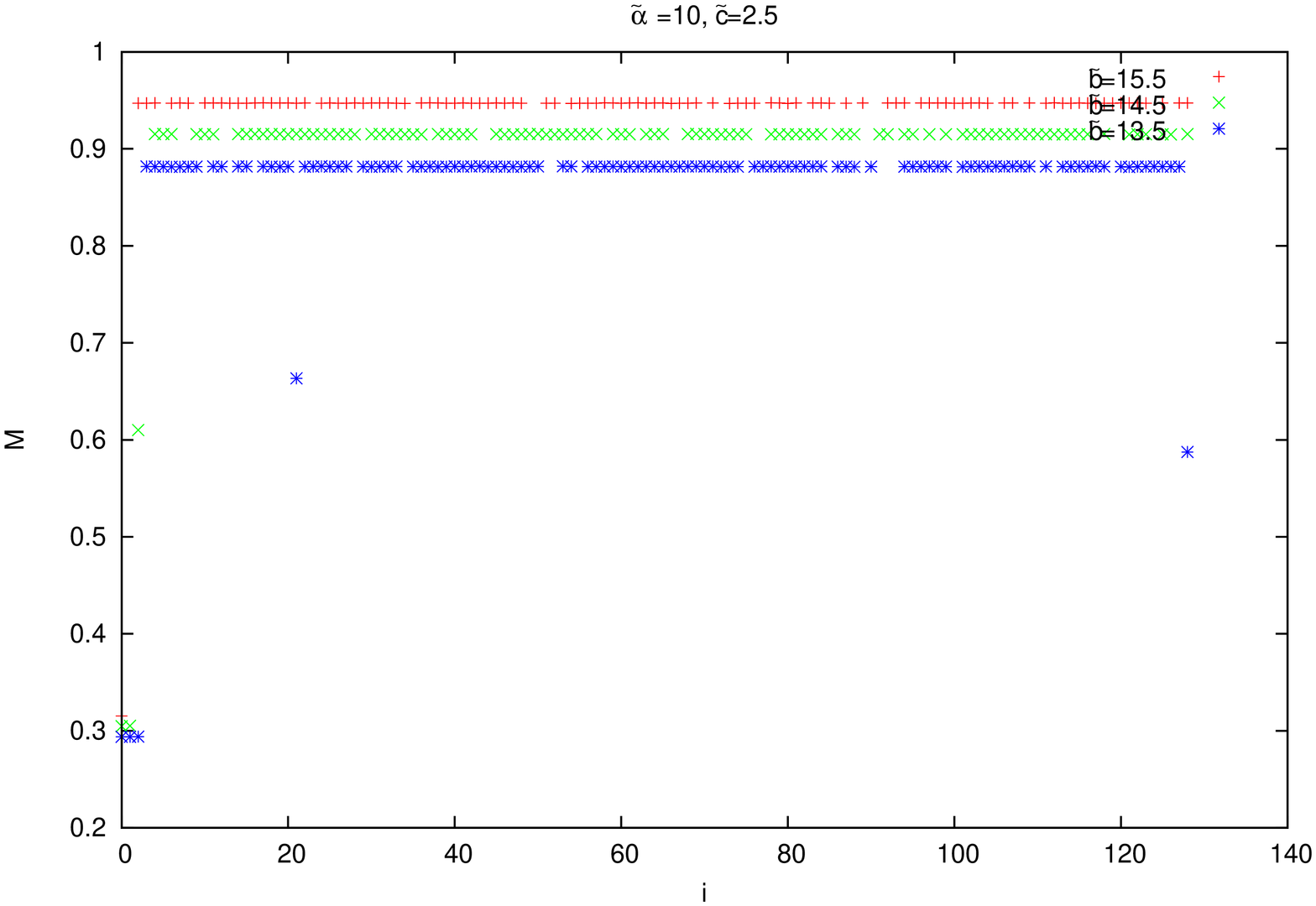}
\includegraphics[width=5.9cm,angle=-90]{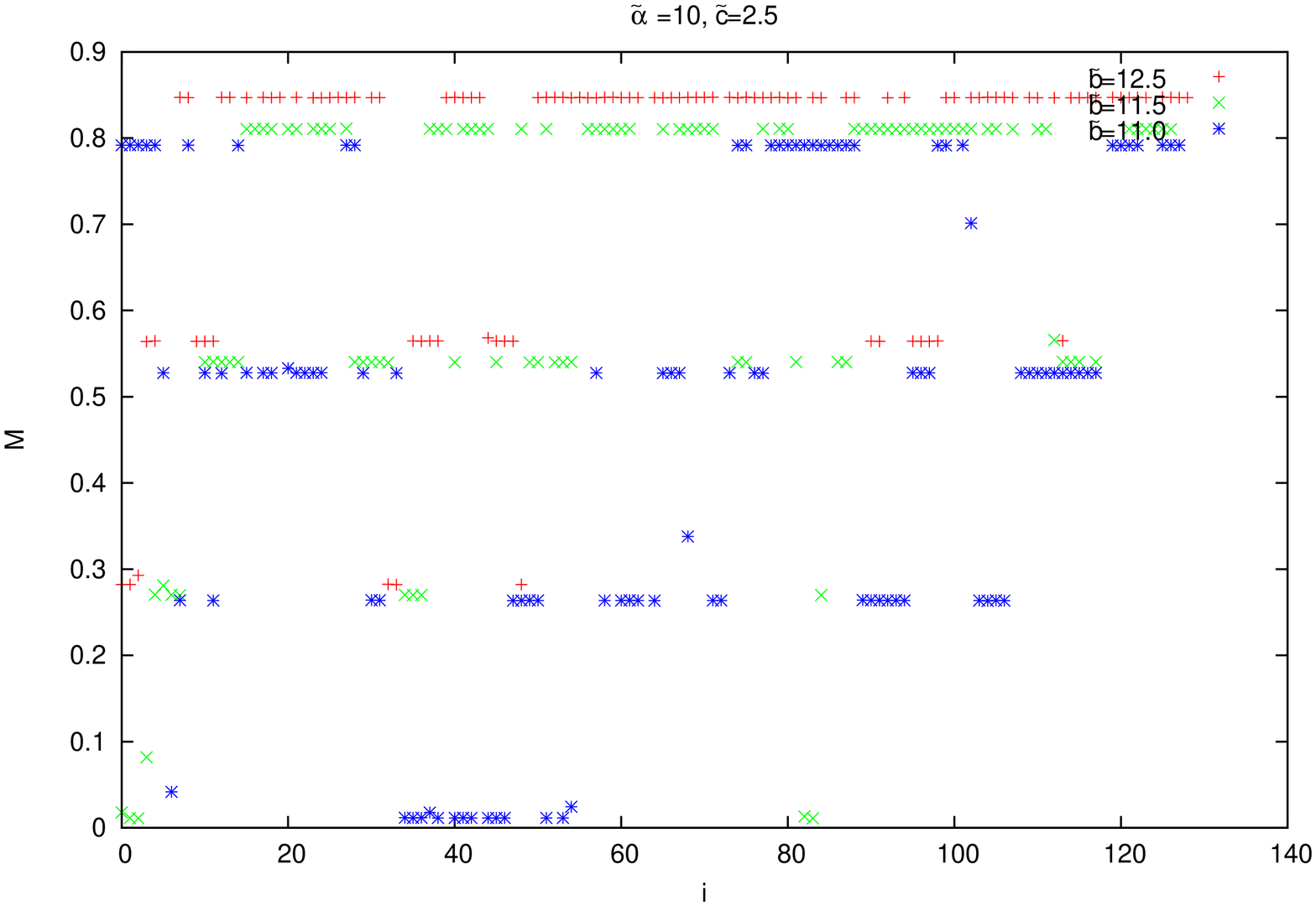}
\includegraphics[width=5.9cm,angle=-90]{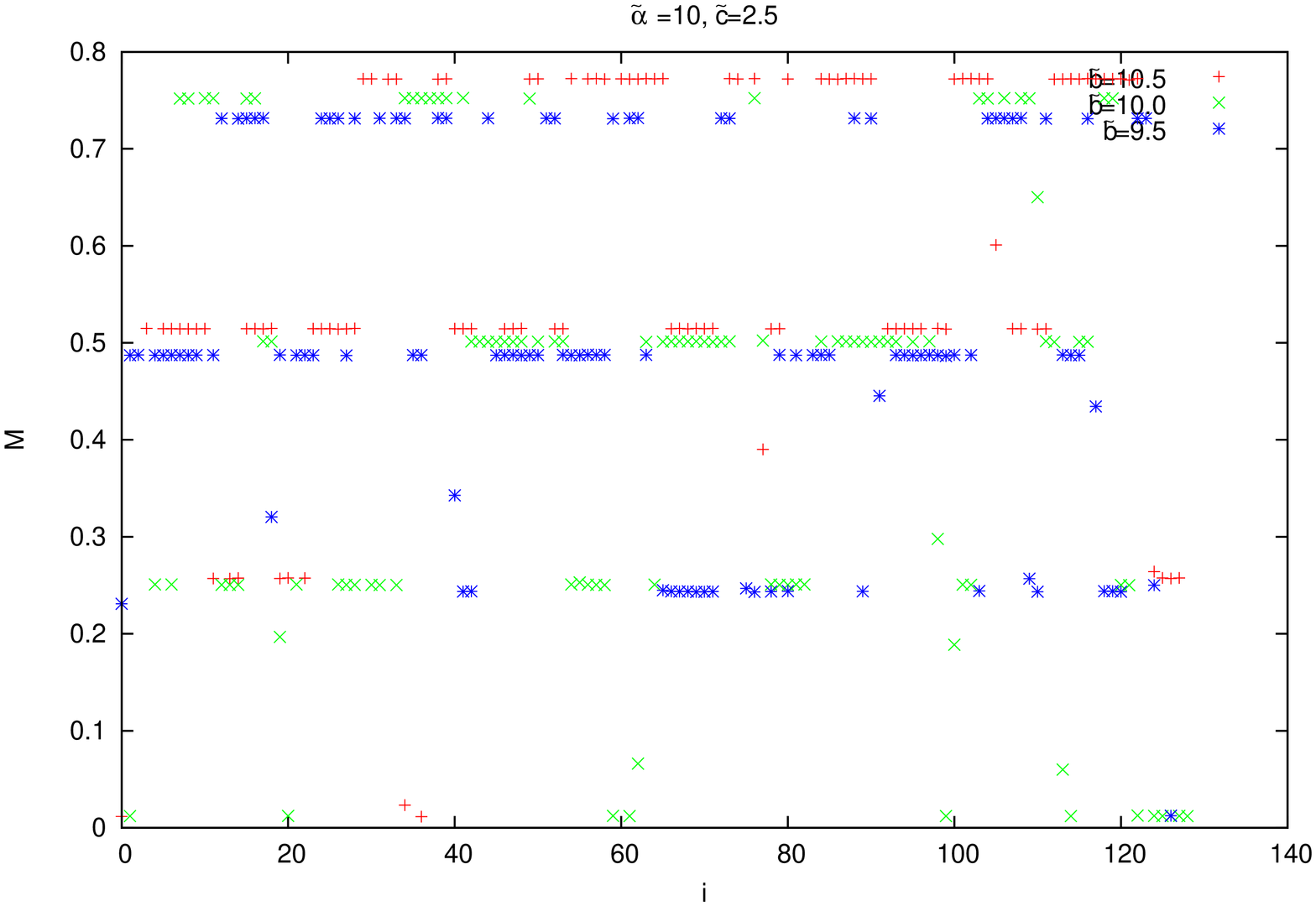}
\includegraphics[width=5.9cm,angle=-90]{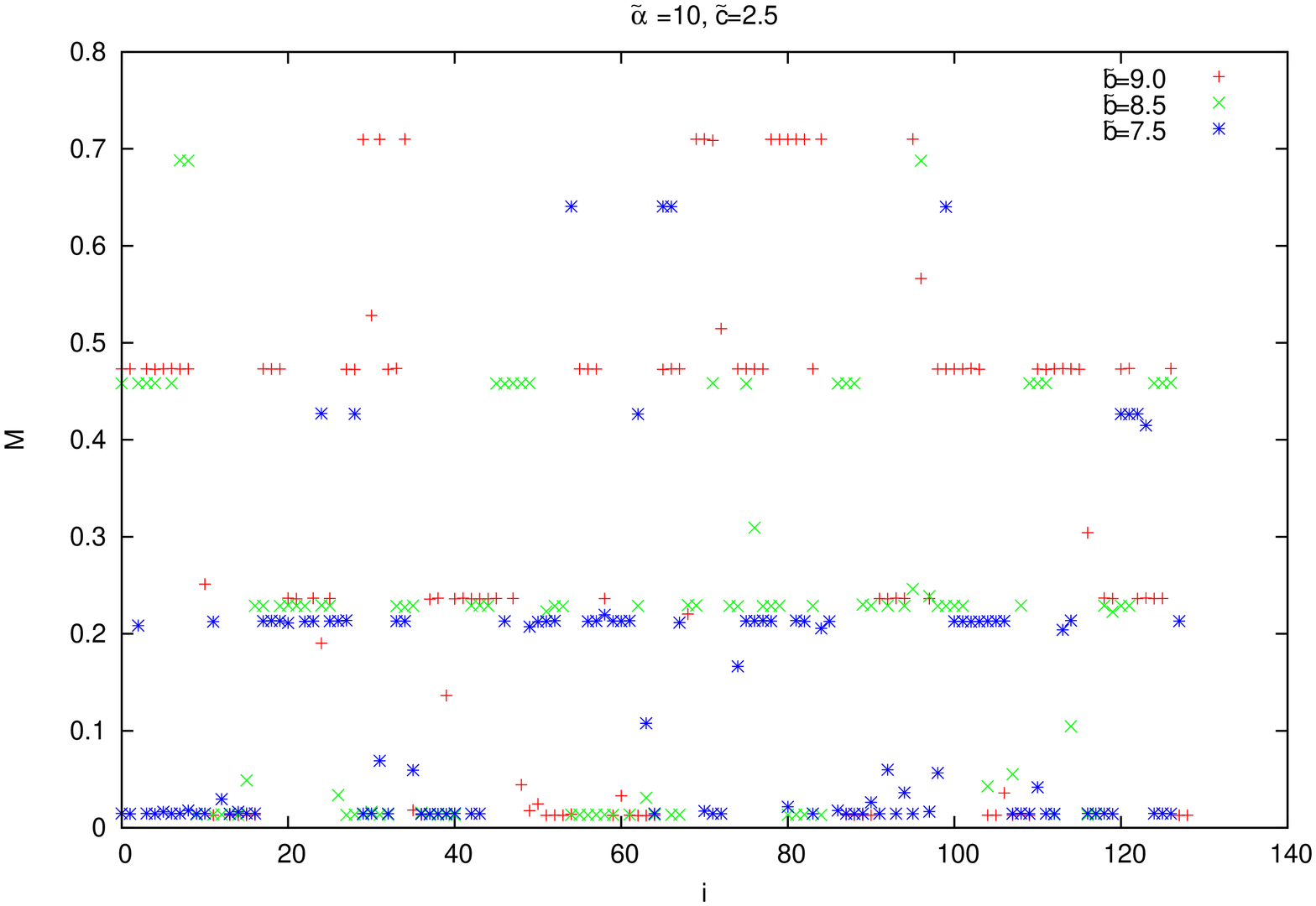}
\caption{Thermalization of the magnetization across the non-uniform-ordered-to-uniform-ordered phase transition. }\label{phase2p0e}
\end{center}
\end{figure}

Some of our results are:
\begin{itemize}

\item{}We present, in figure (\ref{phase2p0}) and (\ref{phase2p0e}), scatter plots for the kinetic action $K$ and the magnetization $M$ respectively, for $\tilde{c}=2.5$, and various values of $\tilde{b}$, for $N=6$. Each point is a single simulation consisting of $2T_{\rm T}+T_{\rm C}\times T_{\rm MC}$ steps. There are at most $T_{\rm S}$ points. The first simulation has been started off from a random $\Lambda$ and  $X_a=\alpha L_a$, whereas each successive simulation is started off from the final state obtained in the previous simulation. 

\item{} We observe that each scatter plot consists of different plateaus, corresponding to the values of the kinetic action/magnetization in the vacua $<\Phi>\sim {\bf 1}$, $\gamma$ and $\gamma_k$. The kinetic action in the vacuum state  $<\Phi>\sim {\bf 1}$ corresponds to  the smallest plateau (almost vanishing), while the kinetic action in the vacuum state  $<\Phi>\sim \gamma$\footnote{We only consider even values of $N$ and thus $\gamma={\rm diag}(+{\bf 1}_{N/2},-{\bf 1}_{\rm N/2})$.} corresponds to the largest plateau. For $N=6$ there are two other vacuum states which are $<\Phi>\sim\gamma_1\sim (+1,+1,+1,+1,+1,-1)$, and $<\Phi>\sim\gamma_2\sim (+1,+1,+1,+1,-1,-1)$, and $\gamma_2$ is approximately degenerate with $\gamma$.

Conversely, the magnetization in the vacuum state  $<\Phi>\sim {\bf 1}$ corresponds to  the largest plateau, while  the magnetization in the vacuum state  $<\Phi>\sim \gamma$ corresponds to the smallest (almost vanishing) plateau. In this case there are clearly four distinct plateaus.

\item{}We observe, in figures (\ref{phase2p0})/(\ref{phase2p0e}), that for large values of $|\tilde{b}|$, thermalized states correspond to  the vacuum states  $<\Phi>\sim {\bf 1}$. See, for examples, the graphs for $\tilde{b}=-15.5,-14.5,-13.5$. These thermalized states are very stable states, and tunneling to other states is very rare, and in fact becomes non-existent as $|\tilde{b}|$ gets larger.

As  $| \tilde{b}|$ decreases, transitions away from  $<\Phi>\sim {\bf 1}$ become more frequent, and scatter plots start showing various plateaus corresponding to the other vacuum states.

As  $| \tilde{b}|$ decreases further, the plateau corresponding to  $<\Phi>\sim {\bf 1}$  becomes virtually empty,  while  the plateaus corresponding to   $<\Phi>\sim \gamma_1$, $<\Phi>\sim \gamma_2$, and $<\Phi>\sim \gamma$   become more populous. For example, for  $\tilde{b}=-9.0,-8.5,-7.5$, it is very rare to see transitions to  $<\Phi>\sim {\bf 1}$, and in fact these transitions become non-existent as $|\tilde{b}|$ gets sufficiently small (but not too small).

\end{itemize}
We conjecture that if we repeat the simulation a sufficient number of times $T_{\rm S}$, then the system will settle into its true minimum. This may take a long time only in the transition region between large 
  and small $|\tilde{b}|$.  It is immediately obvious, from the above discussion, that for large  $|\tilde{b}|$ the minimum is  $<\Phi>\sim {\bf 1}$, while for small  $|\tilde{b}|$  the minimum is  $<\Phi>\sim \gamma$.

\paragraph{Eigenvalues Distributions:}
It is quite obvious, that the most revealing order parameter, is the eigenvalue distribution of the scalar field $\Phi$. In our approach, the eigenvalues are precisely the degrees of freedom which we are sampling. We  can then use immediately  the  $T_{\rm MC}$ sets of eigenvalues $\lambda_i$ obtained in the Monte Carlo evolution, for a fixed $\tilde{c}$ and $\tilde{b}$, to 
 construct  appropriate histograms. These are precisely the eigenvalue distributions $\rho(\lambda)$ of the scalar field $\Phi$. 

\begin{figure}[htbp]
\begin{center}
\includegraphics[width=5.9cm,angle=-90]{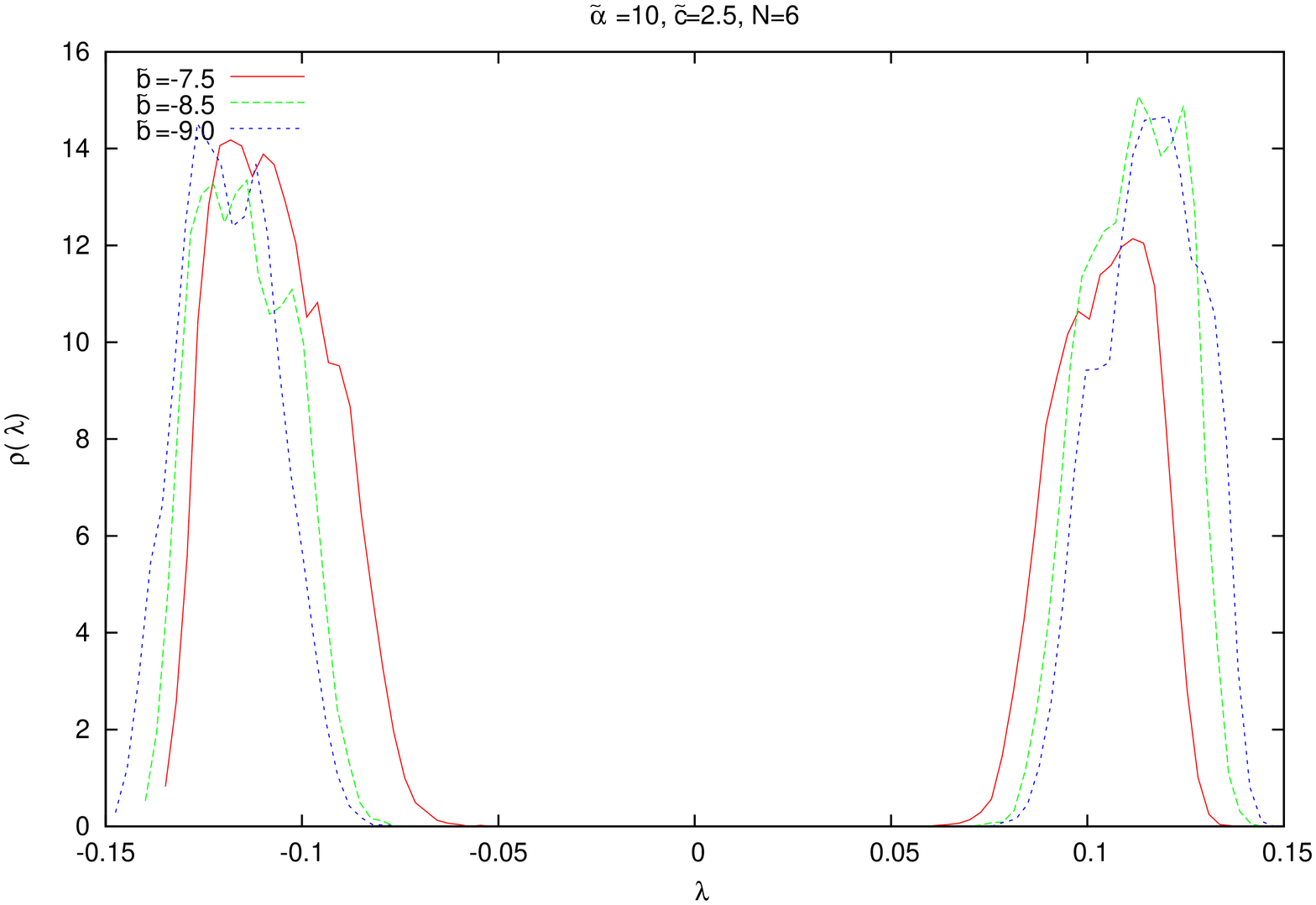}
\includegraphics[width=5.9cm,angle=-90]{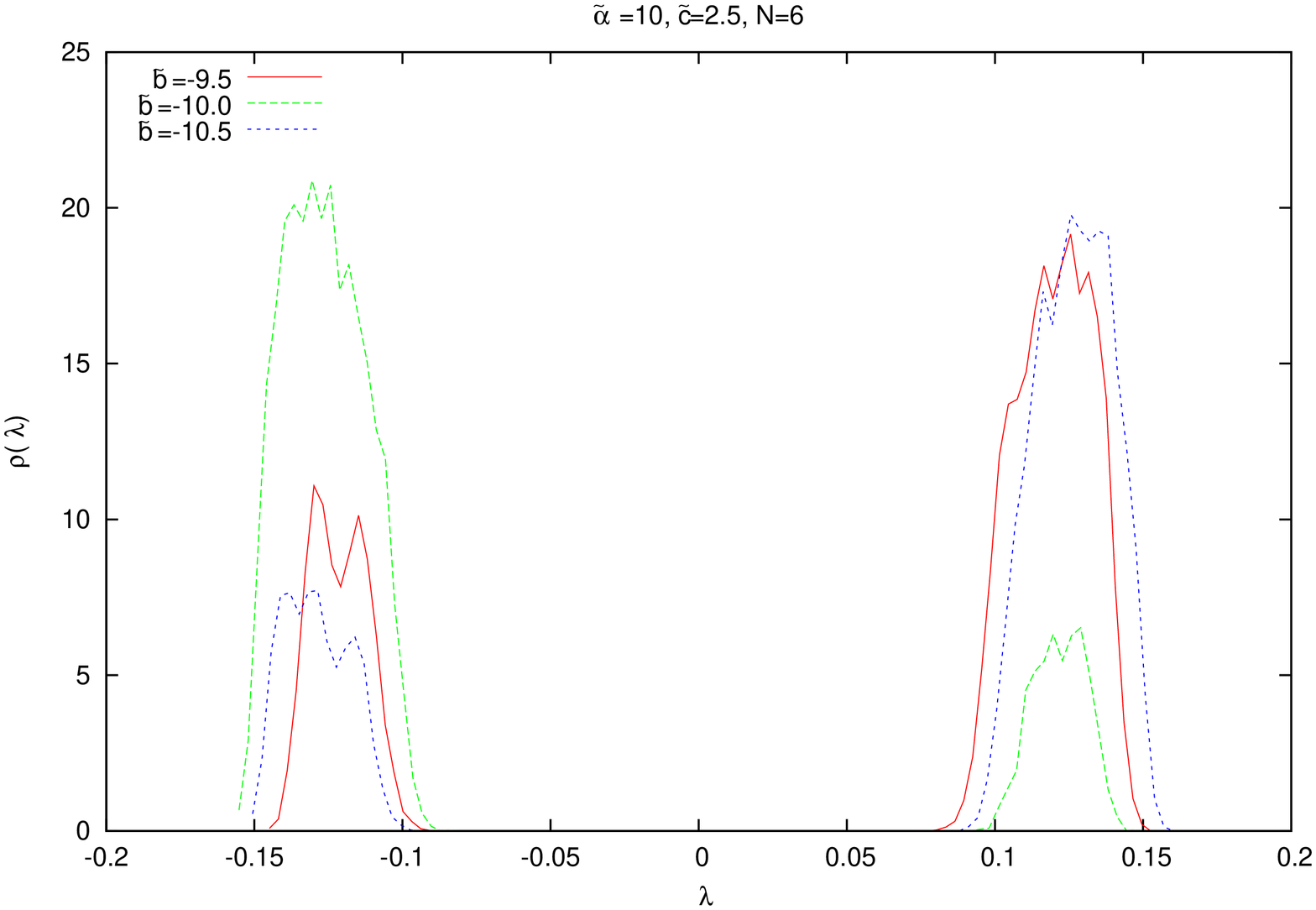}
\includegraphics[width=5.9cm,angle=-90]{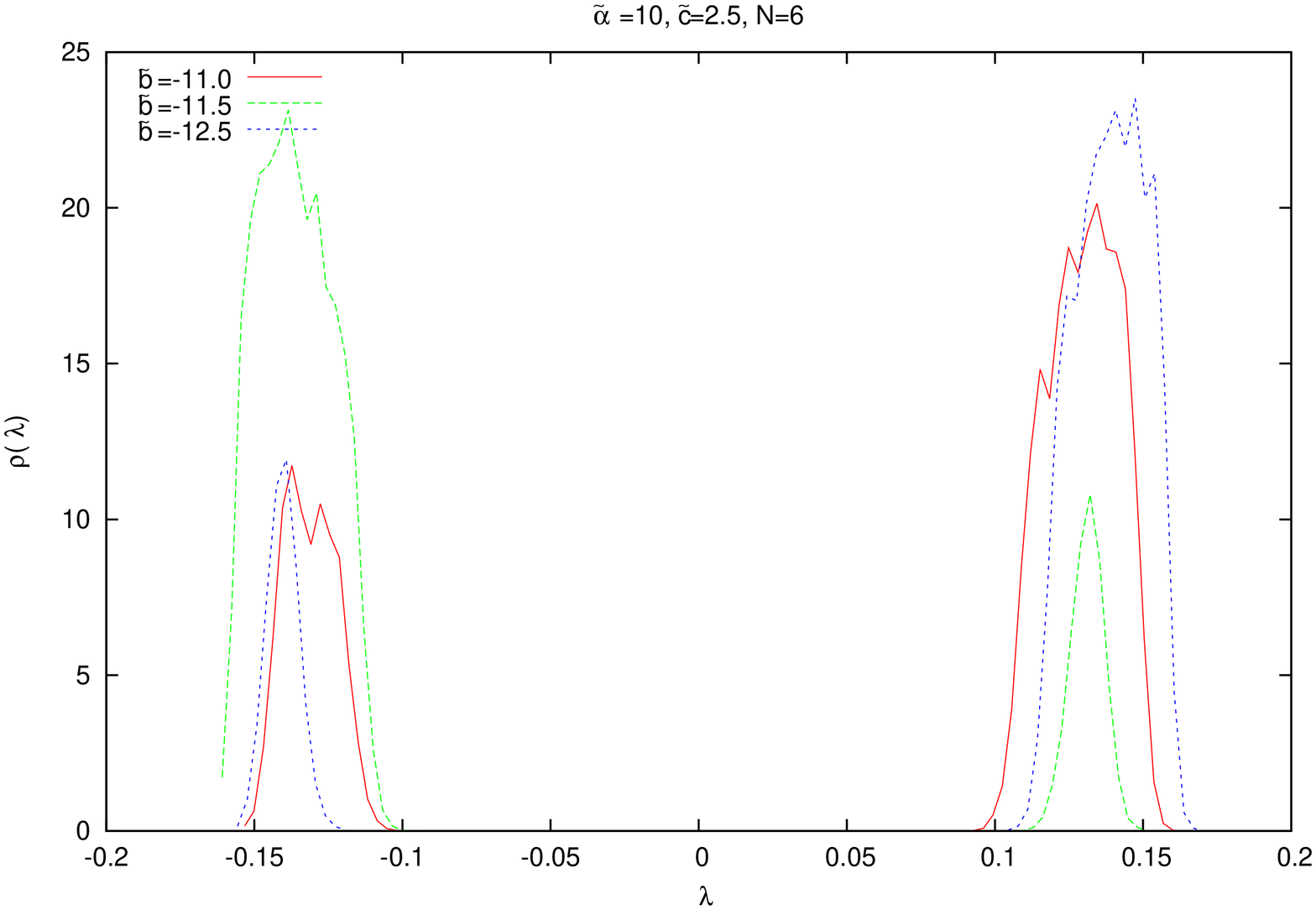}
\includegraphics[width=5.9cm,angle=-90]{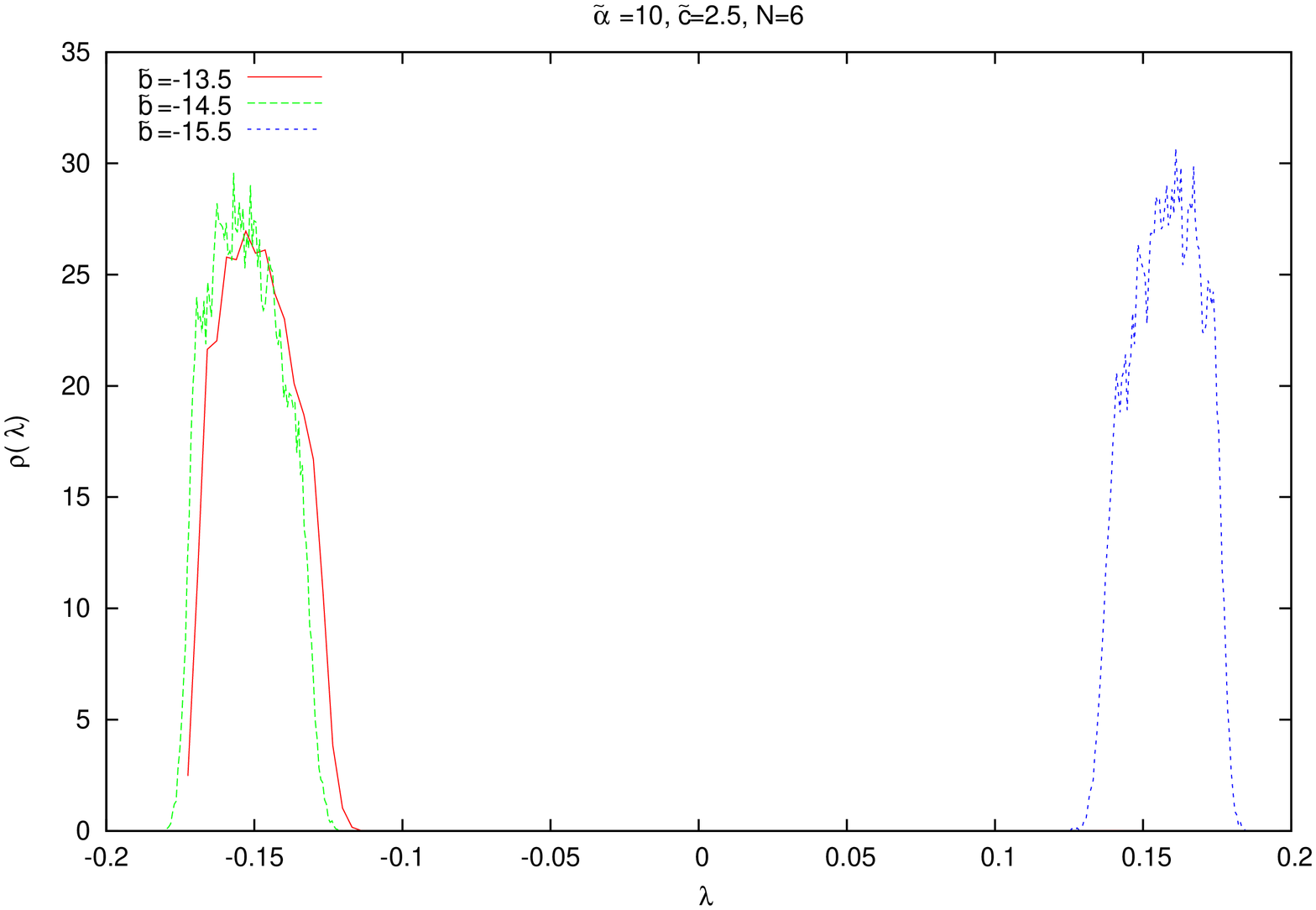}
\caption{The eigenvalue distributions across the uniform-to-non-uniform transition. }\label{ev_NU_U}
\end{center}
\end{figure}

In figure (\ref{ev_NU_U}), we plot the eigenvalue distributions for various values of $\tilde{b}$, across the uniform-to-non-uniform transition point, for $N=6$ and $\tilde{c}=2.5$. We observe that we go from the one-cut solution, centered about $+\sqrt{-r/2u}$, to the two-cut solution, centered    about $\pm \sqrt{-r/2u}$, around $\tilde{b}=-10.5\pm 0.5$, which agrees with our other measurement (see below). 

Although in the two-cut solution we know that the eigenvalues are  $\pm \sqrt{-r/2u}$, we can not tell how many of them are pluses, and how many of them are minuses\footnote{The order of the pluses and minuses is irrelevant, i.e. it can not be observed.}. In order to determine the distribution of the plus and minus signs, we may then plot, the probability distribution of the values of the magnetization $Tr \Phi$. Alternatively, we can directly look at the eigenvalues themselves, to see which matrices are involved. As it turns out, in the transition region between large and small $|\tilde{b}|$, the vacuum states are not given simply by the pure states ${\bf 1}$, $\gamma$, $\gamma_1$ and $\gamma_2$ ,but they are, typically, given by admixture of these pure states. 


\paragraph{Critical Values:}

According to  \cite{GarciaFlores:2009hf}, the  non-uniform-ordered-to-uniform-ordered transition, should occur at the value of $\tilde{b}$, where the susceptibility and the specific heat are peaked, which is something we were not able to reproduce in our scheme in any consistent way.

The determination of the location of the  non-uniform-ordered-to-uniform-ordered  transition, can also be based, on the location of  the "discontinuity/jump" in the expectation value of the kinetic term.  This discontinuity is also associated with a discontinuity in the total power, power in the zero mode and magnetization. 

As opposed to all other simulations reported in this article, we will attempt  in the current case to cross the critical line by holding $-\tilde{b}$ fixed, while varying $\tilde{c}$. In this way, we are guaranteed to cross, first, the  non-uniform-ordered-to-uniform-ordered transition, as we increase $\tilde{c}$, at some fixed value of $-\tilde{b}$. If we fix $\tilde{c}$ instead, and start increasing $-\tilde{b}$, we will hit the matrix phase transition first (see next subsection), then the  non-uniform-ordered-to-uniform-ordered transition.

The detail of this simulation goes as follows. The initial state, for a fixed $\tilde{b}$ and $\tilde{c}$, is prepared by starting from a random configuration $\Lambda$, and from $X_a=\alpha L_a$, and then run a Metropolis updating procedure for $T_{\rm T}$ steps on this initial state,  at fixed $X_a$ without taking into account the effect of the Vandermonde determinant. We repeat this process for $T_S=2^{4}$ steps to get the actual initial state.   Starting from this resulting state, we run a  full Metropolis updating procedure for  $T_{\rm T}+T_{\rm C}\times T_{\rm MC}$. 

\begin{figure}[htbp]
\begin{center}
\includegraphics[width=5.9cm,angle=-90]{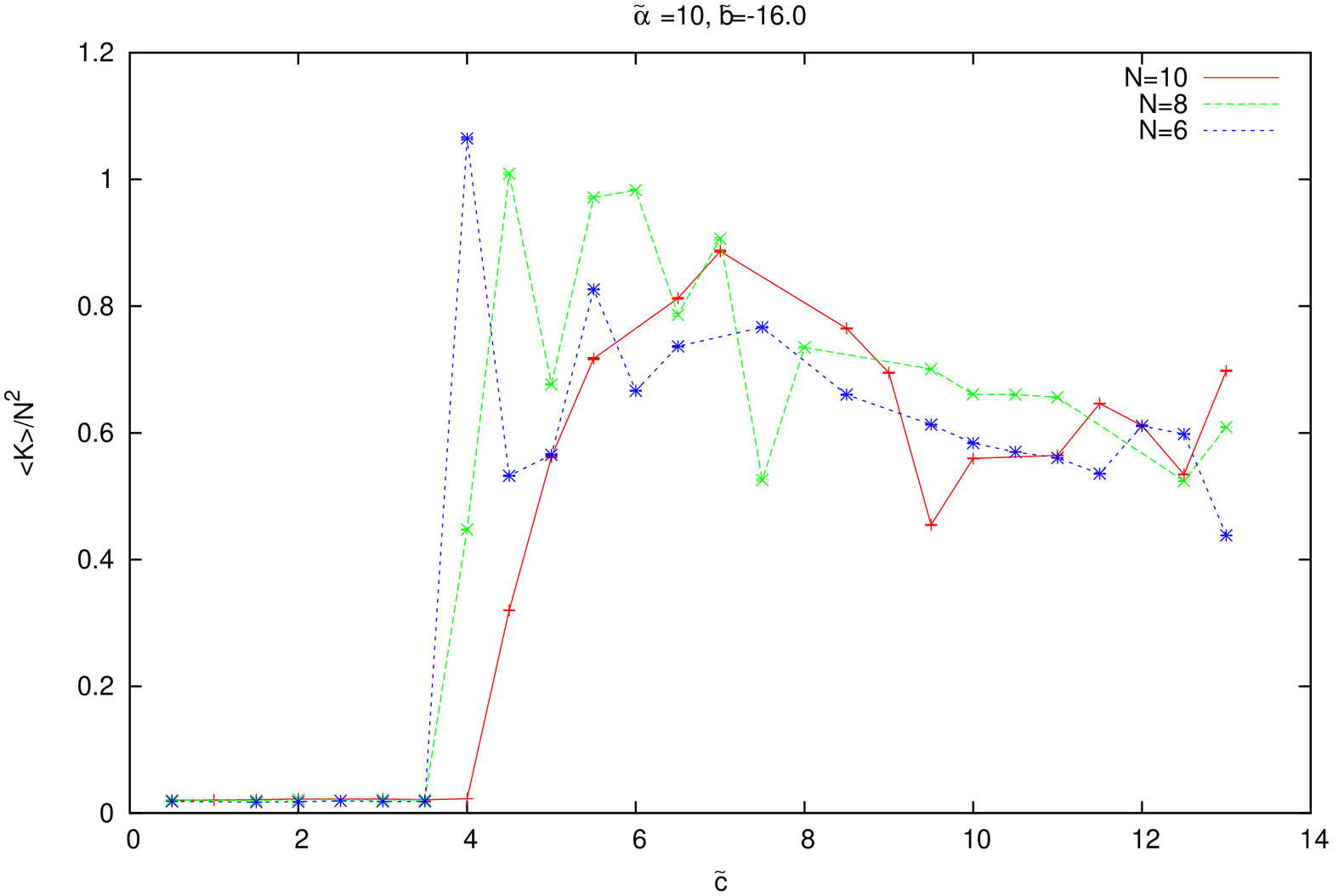}
\includegraphics[width=5.9cm,angle=-90]{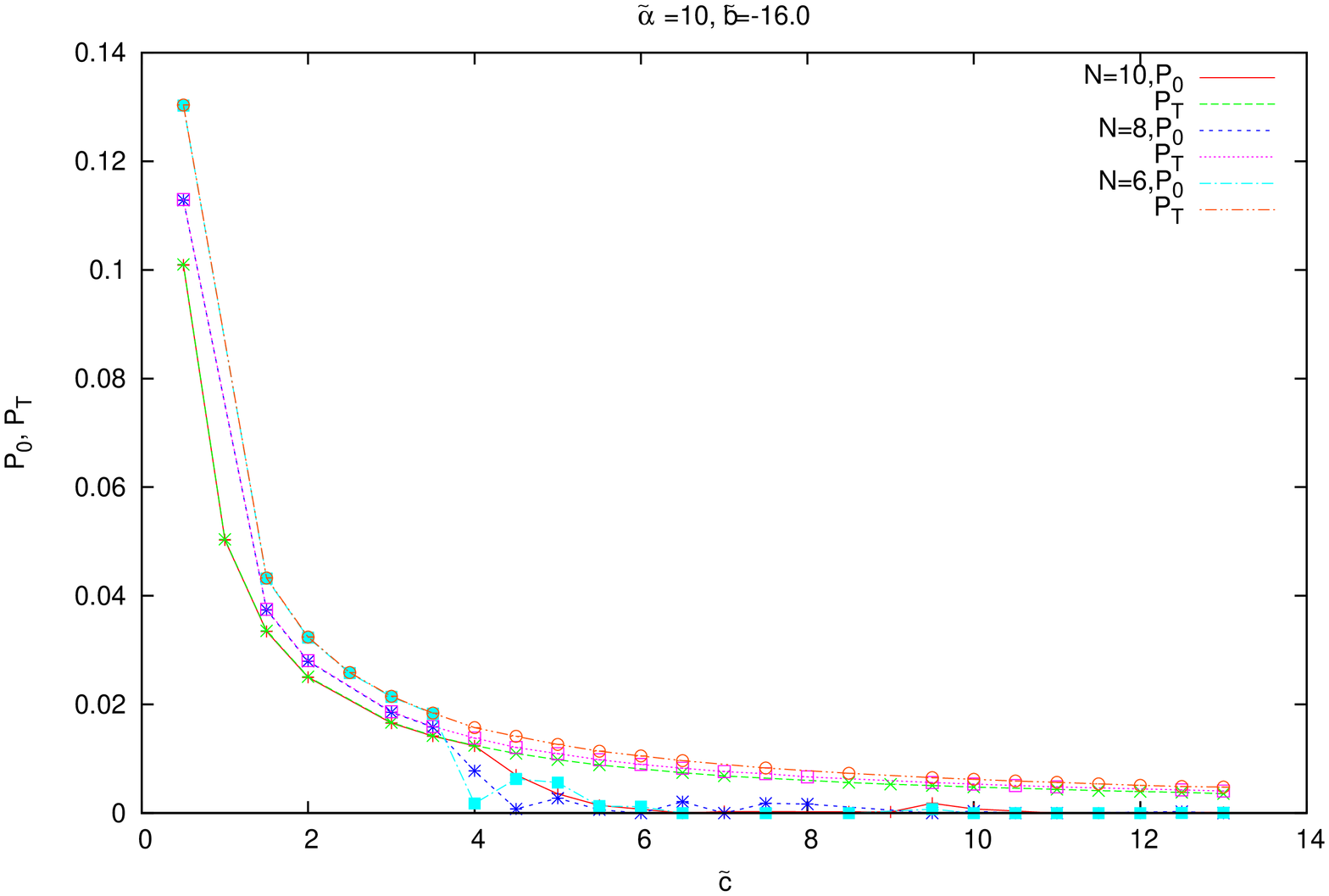}
\includegraphics[width=5.9cm,angle=-90]{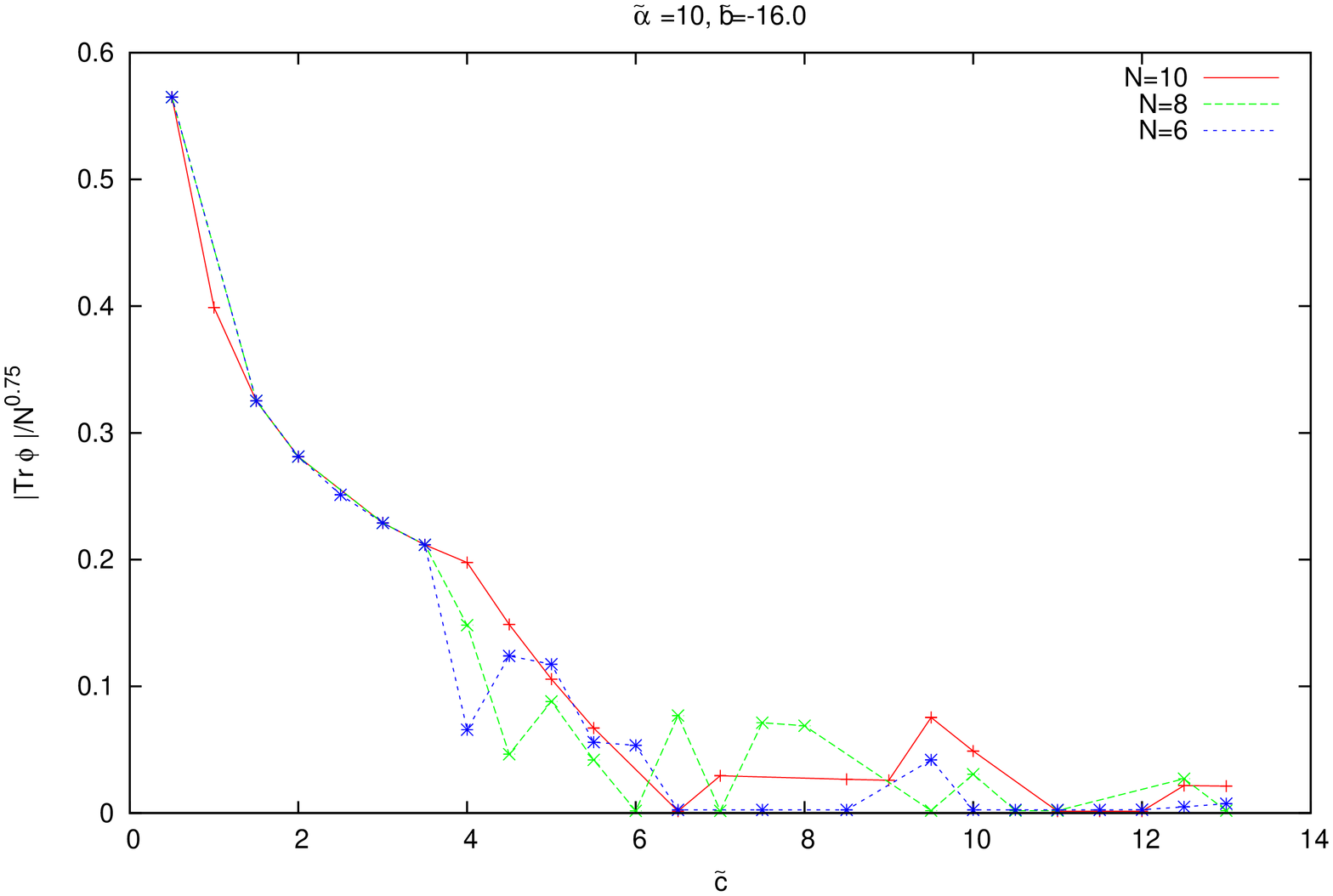}
\caption{The non-uniform-ordered-to-uniform-ordered phase transition. }\label{phase2p2}
\end{center}
\end{figure}

We work always with $T_{\rm T}=T_{\rm MC}=2^{13}$, and $T_{\rm C}=2^4$, for $N=6,8,10$. The constraint on the identity is $<{\rm IDE}>/N^2=4.00\pm 0.30$ ($N=6$), and $<{\rm IDE}>/N^2=4.00\pm 0.25$ ($N=8,10$). The results are shown on figure (\ref{phase2p2}). In the graphs of the total power, and the power in the zero mode, we can find from the scaling (\ref{scaling}), that in the Ising phase $P_0=P_T\sim - \tilde{b}/(\sqrt{N}\tilde{c})$, which is why the graphs for the powers for different $N$ do not collapse.

We will take, as our measurement of the  non-uniform-ordered-to-uniform-ordered transition points, the arithmetic average of the critical points, obtained  from the discontinuity/jump in the expectation value of the kinetic term for different $N$\footnote{The underlying assumption here is that the measurements for different $N$ are, actually, the same and differences between them are only due to the limitation of the simulations.}. We drop here the calculation of the error bars which requires much more efforts. Some results are given in  table (\ref{table0}). 

The fit to the  non-uniform-ordered-to-uniform-ordered transition line, as computed from table (\ref{table0}),  is given by $\tilde{c}=-0.22\tilde{b}+0.38$. The slope is very close to the slope of the Ising transition line given by equation (\ref{ising}). This confirms the general conjecture of   \cite{GarciaFlores:2009hf}, that the  non-uniform-ordered-to-uniform-ordered transition line, is the continuation, of the Ising transition line, to general values of $\tilde{c}$ and $\tilde{b}$. However, the intercept of the fit $\tilde{c}=-0.22\tilde{b}+0.38$ seems to be quite large. We claim that this is, only, due to our limited number of data points, 
 and lack of error bars. Clearly, for $\tilde{c}=0$, there is no Ising transition, nor a non-uniform-ordered-to-uniform-ordered transition. In other words,   the  non-uniform-ordered-to-uniform-ordered transition line must  go  through the origin $(\tilde{c},\tilde{b})=(0,0)$. The fit to the  non-uniform-ordered-to-uniform-ordered transition line, as computed from table (\ref{table0}) plus the point $(\tilde{c},\tilde{b})=(0,0)$,  is now given by
 \begin{eqnarray}
\tilde{c}=-0.25\tilde{b}+0.03.\label{NU-U}
\end{eqnarray}
The error in the intercept is found to be $0.1$, while the error bar in the slope is negligible. The measured slope, as well as the measured small intercept, are reasonably close to the values measured in  \cite{GarciaFlores:2009hf}.

\begin{table}[h]
\centering
\begin{tabular}{|l|c|c|c|c|c| }
\hline
$\tilde{b}$& $N=10$  & $N=8$ & $N=6$ &  $\tilde{c}_* ({\rm arithmetic}~{\rm average})$ & $\tilde{c}({\rm FDX})$\\
\hline 
$-10.0$ & $3.25$ &  $2.25 $ & $2.25  $ & $2.58 $ &  $2.07$\\ 
$-16.0$ & $4.25$  & $3.75$ & $3.75 $ & $3.92 $ &  $3.27$\\ 
\hline 
\end{tabular}
\caption{The  non-uniform-ordered-to-uniform-ordered transition points. }\label{table0}
\end{table}


\subsection{The Matrix Phase Transition}  
The non-uniform-ordered-to-disordered transition, also called matrix transition,    appears for medium and large values of $\tilde{c}$. 
We perform simulations in a similar fashion to the Ising case, with the exception that we start from a random configuration for each value of $\tilde{b}$. We take $\tilde{b}$ in the range $[-15,0]$, with step equal $0.25$, and values of $\tilde{c}$ in the range $[2,25]$. 



\paragraph{The Matrix Transition in the Limit of Large Couplings:}

It is expected that for large values of the coupling constant $\tilde{c}$, the matrix transition in the full model, will be given approximately, by the matrix transition in the pure potential model, i.e. the model without kinetic term. This approach becomes exact in the limit $\tilde{c}\longrightarrow \infty$. 

\begin{figure}[htbp]
\begin{center}
\includegraphics[width=5.9cm,angle=-90]{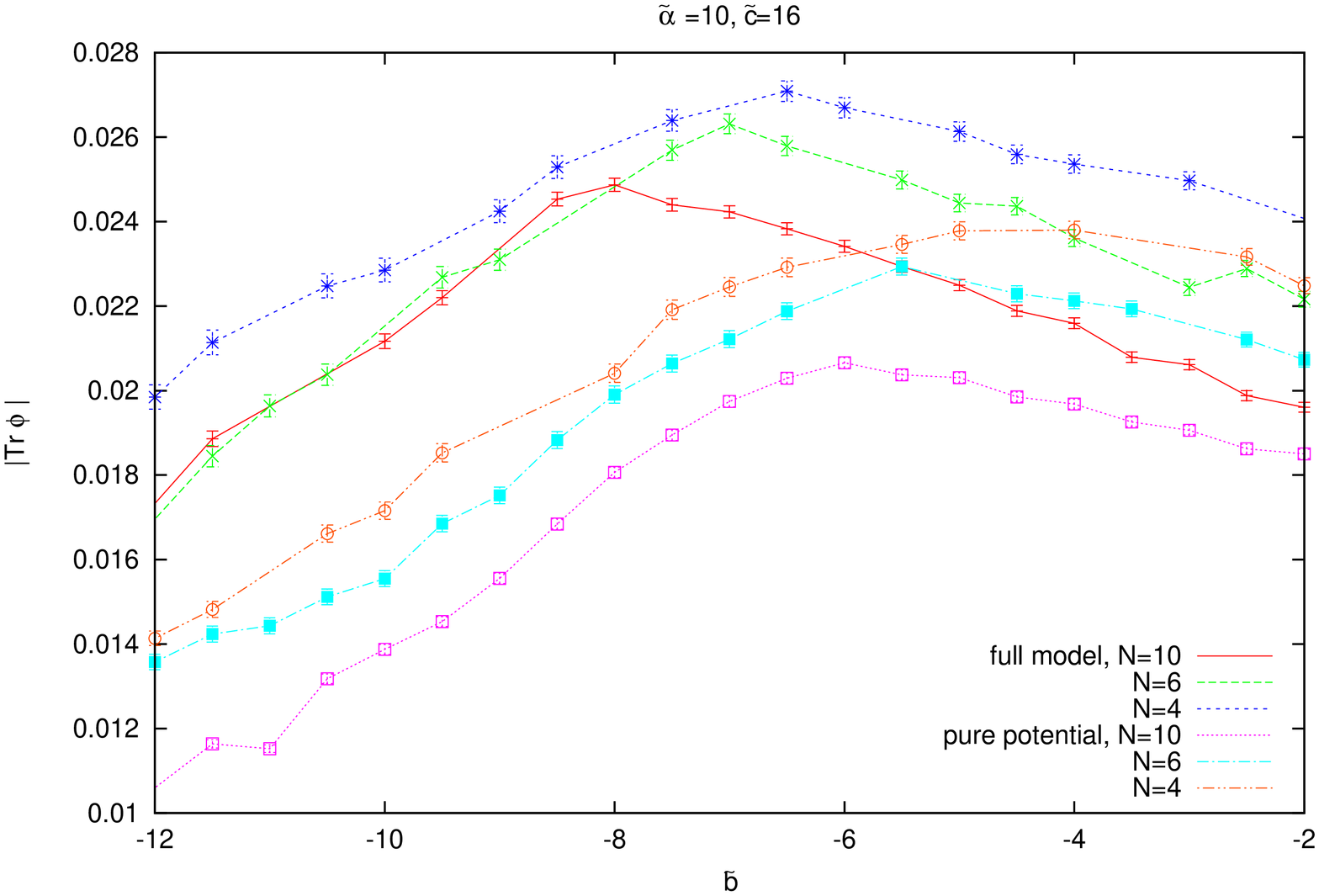}
\includegraphics[width=5.9cm,angle=-90]{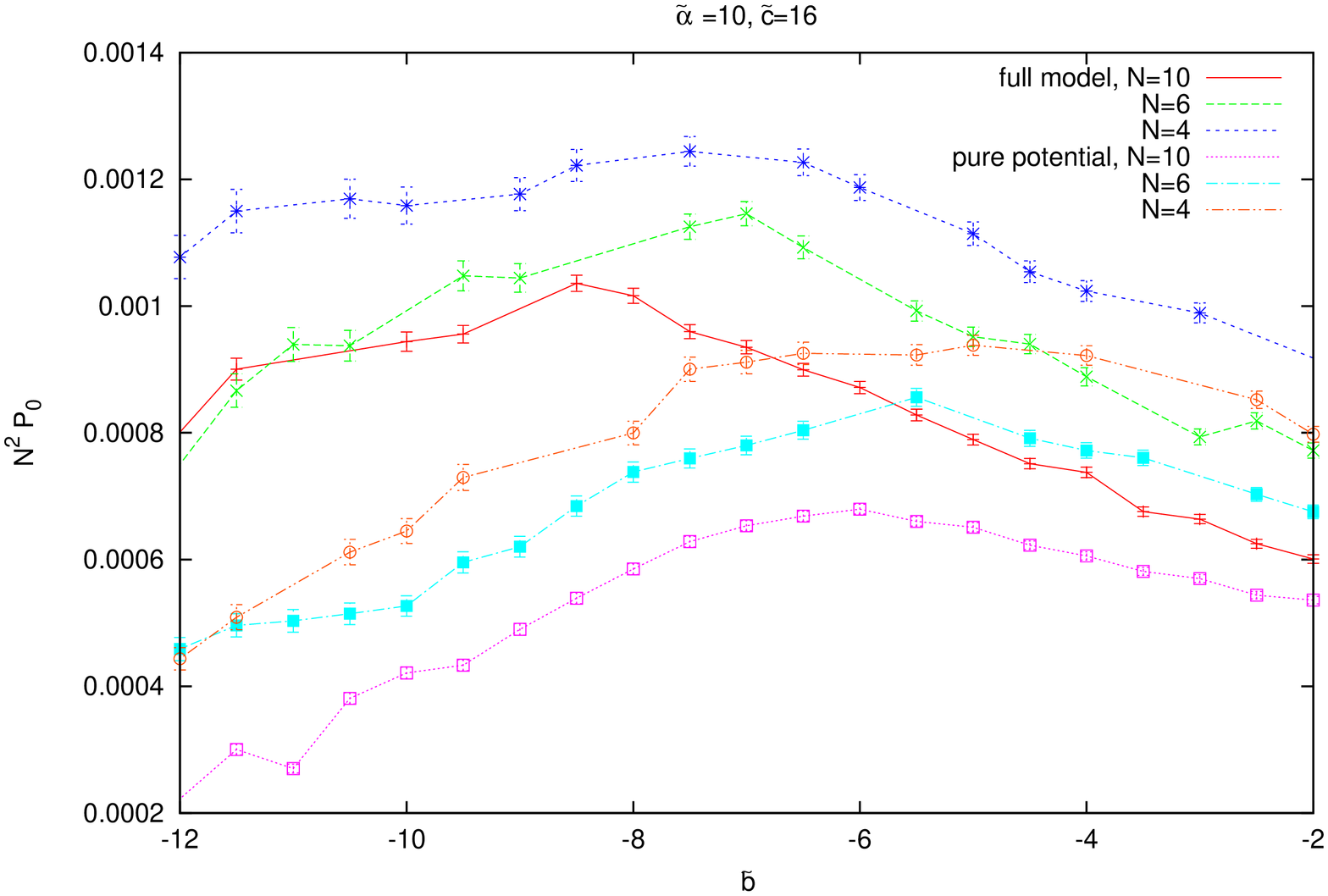}
\includegraphics[width=5.9cm,angle=-90]{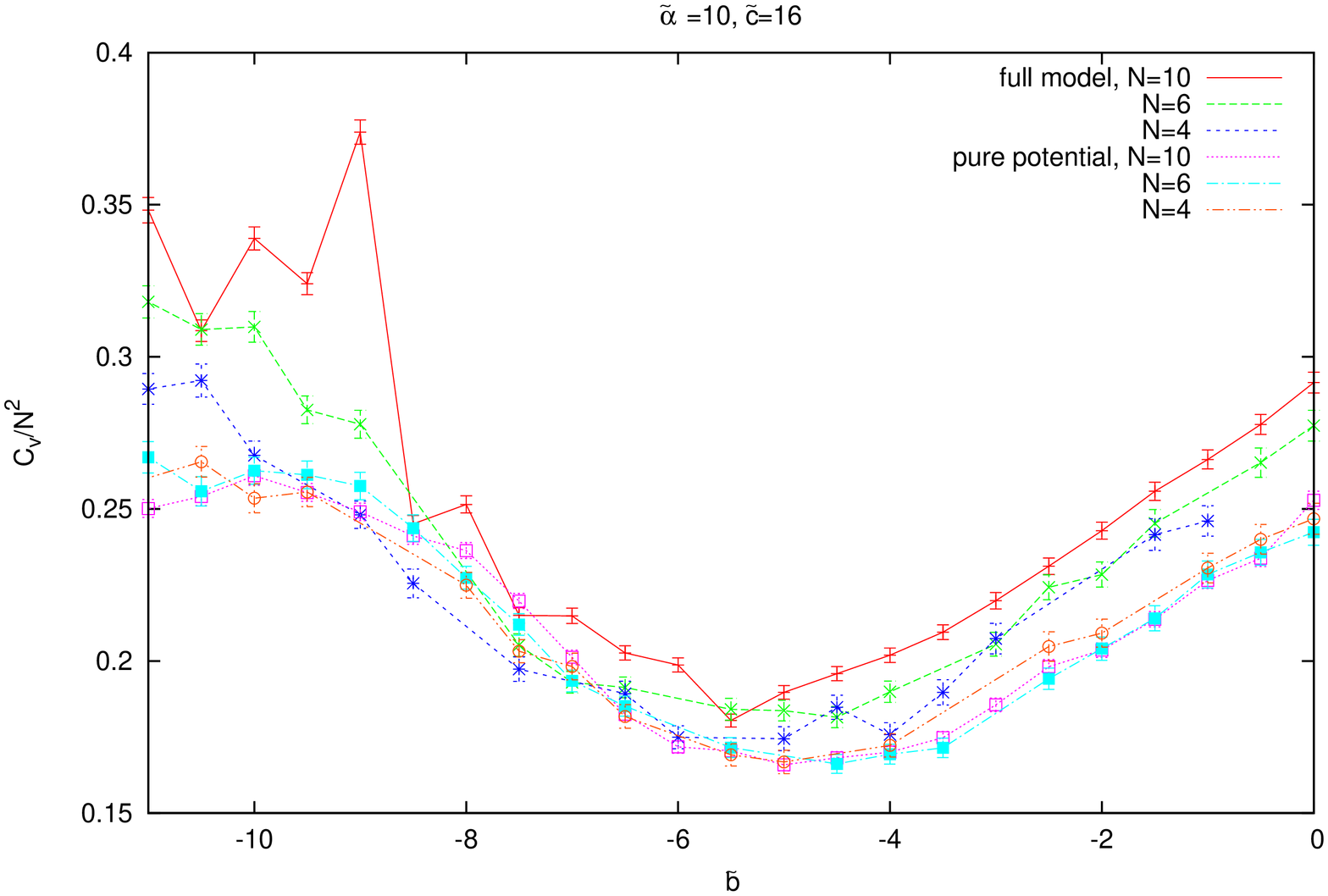}
\includegraphics[width=5.9cm,angle=-90]{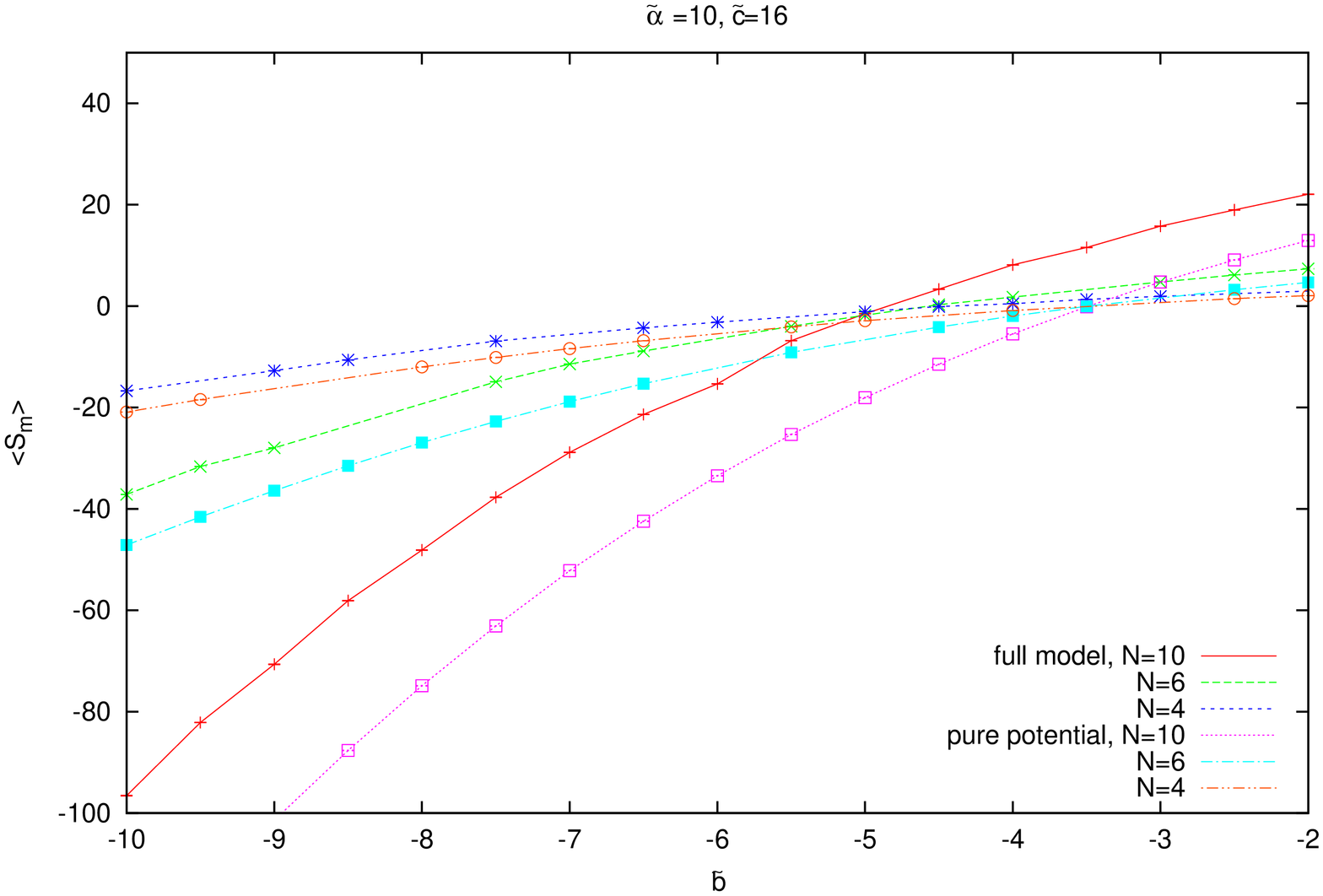}
\caption{The matrix phase transition.
}\label{phase3}.
\end{center}
\end{figure}

We include in figure (\ref{phase3}), the behavior of the magnetization $M=|{\rm Tr}\Phi|$, the zero power (power in the zero modes) $N^2P_0$, the sepcific heat $C_v/N^2$, and the average action $<S_m>$ for $\tilde{c}=16$.  We plot the pure potential model for comparison. 

It is well known that the matrix transition occurs, in the pure potential model, at the point where the specific heat divided by the number of degrees of freedom becomes equal to $1/4$, after passing through its minimum as we increase $|\tilde{b}|$. This corresponds, for any fixed value $\tilde{c}$,  to the transition point $\tilde{b}_*=-2\sqrt{\tilde{c}}$. 

This transition is anticipated by the intersection point, which is $N-$independent, seen on the graph of the action $<S_m>$, and by the location of the wide maximum, seen on the graphs of the magnetization $M$ and the zero power $P_0$. However, all these estimates, provide only an  under estimation of the actual transition point in the pure potential model.

If we take, as our measurement of the matrix transition in the full model, the point where the specific heat becomes equal to $1/4$ after passing through its minimum, then we find, as opposed to the pure potential model, an under estimation of the transition point. The intersection point of the action $<S_m>$ provides, as 
before, also an under estimation of the transition point.



In the full model, we have observed that, for sufficiently large values of $\tilde{c}$,  a reasonable estimation of the matrix transition point, which compares favorably to the theoretical prediction coming from the pure potential model,   can be given by the location of the broad maximum, seen on the graphs of the magnetization and the zero power. 

We search for this maximum for values of $\tilde{b}$ much smaller than the discontinuity point relevant for the non-uniform-to-uniform transition. 

We include  in table (\ref{table1}), our measurements of the matrix transition point, for $N=4,6,10$, coming from the magnetization (first measurement), and the zero power (second measurement), and compare them with the pure potential model prediction.

\begin{table}[h]
\centering
\begin{tabular}{|l|c|c|c|c|c|c| }
\hline
$\tilde{c}$ &  $N=4$ &  $N=6$ & $N=10$ &  $\tilde{b}_* ({\rm arithmetic}~{\rm average})$ & $\tilde{b}_* ({\rm pure}~{\rm matrix}~{\rm model},$ \\
 &   &   &  &   & ${\rm theory})$ \\
\hline 
$25.0$ &  $-7.33\pm 1.33$& $-9.25\pm 0.75$ & $-9\pm 0.5$ & $-9.07\pm 1.07$ & $-10$ \\
 &   $-8.67\pm 2.17$&  $-9.67\pm 1.17$ &  $-10.5\pm 0.5$ & &  \\
$16.0$ &  $-6.67\pm 0.67$ & $-7\pm 0.5$ & $-8\pm 0.5$ & $-7.47\pm 0.64$ & $-8$ \\
 &  $-7.5\pm 1$ & $-7\pm 0.5$ & $-8.67\pm 0.67$ &  &  \\
$9.0$ &  $-5.33\pm 0.83$& $-5.33\pm 0.83$ & $-6.5\pm 1$ & $-6.11\pm 0.94$ & $-6$ \\
 &  $-6.67\pm 0.67$& $-6.33\pm 1.33$ & $-6.5\pm 1$ &  &  \\
\hline 
\end{tabular}
\caption{The matrix transition points for the full model for large couplings. }\label{table1}
\end{table}

\begin{figure}[htbp]
\begin{center}
\includegraphics[width=5.9cm,angle=-90]{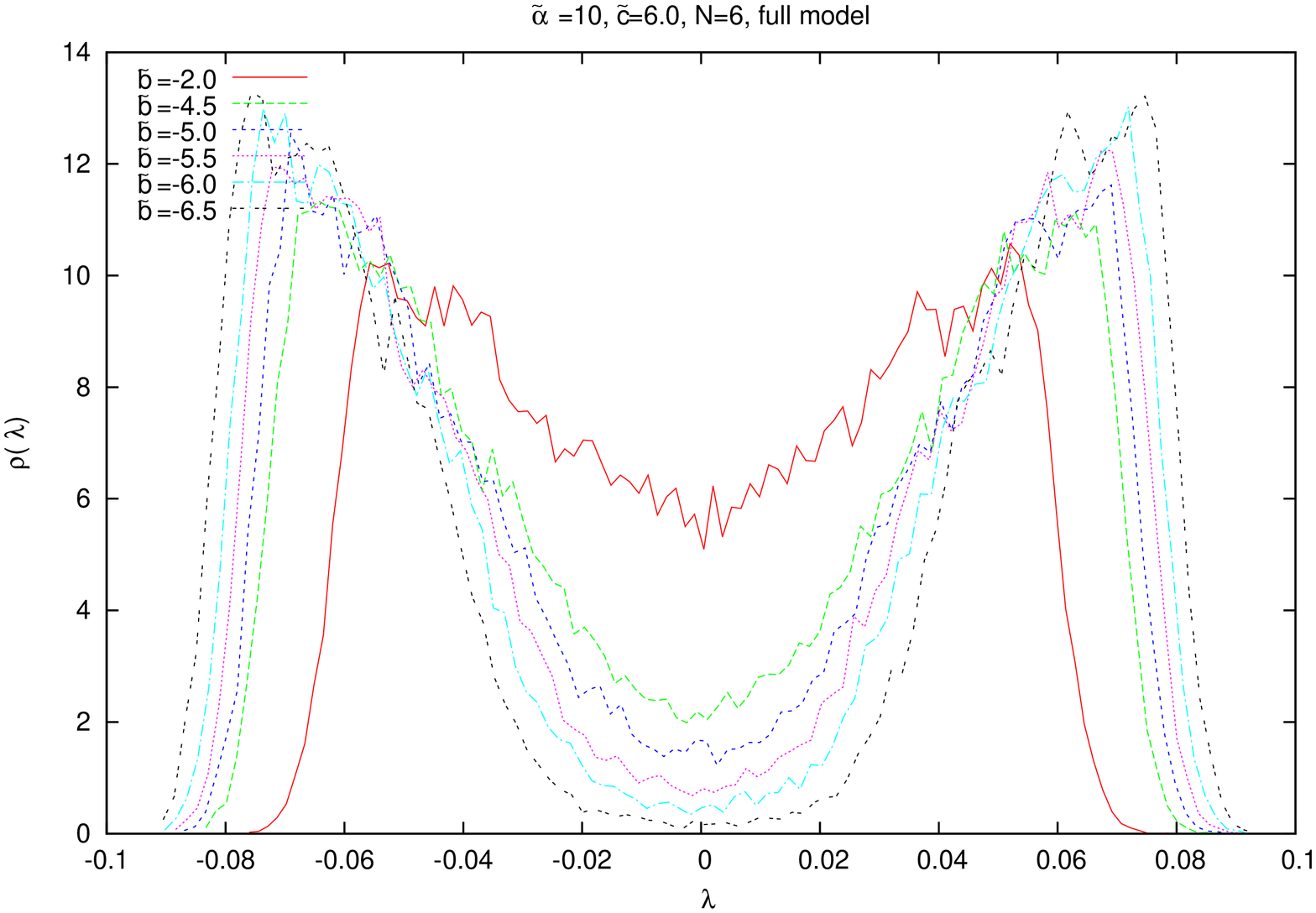}
\includegraphics[width=5.9cm,angle=-90]{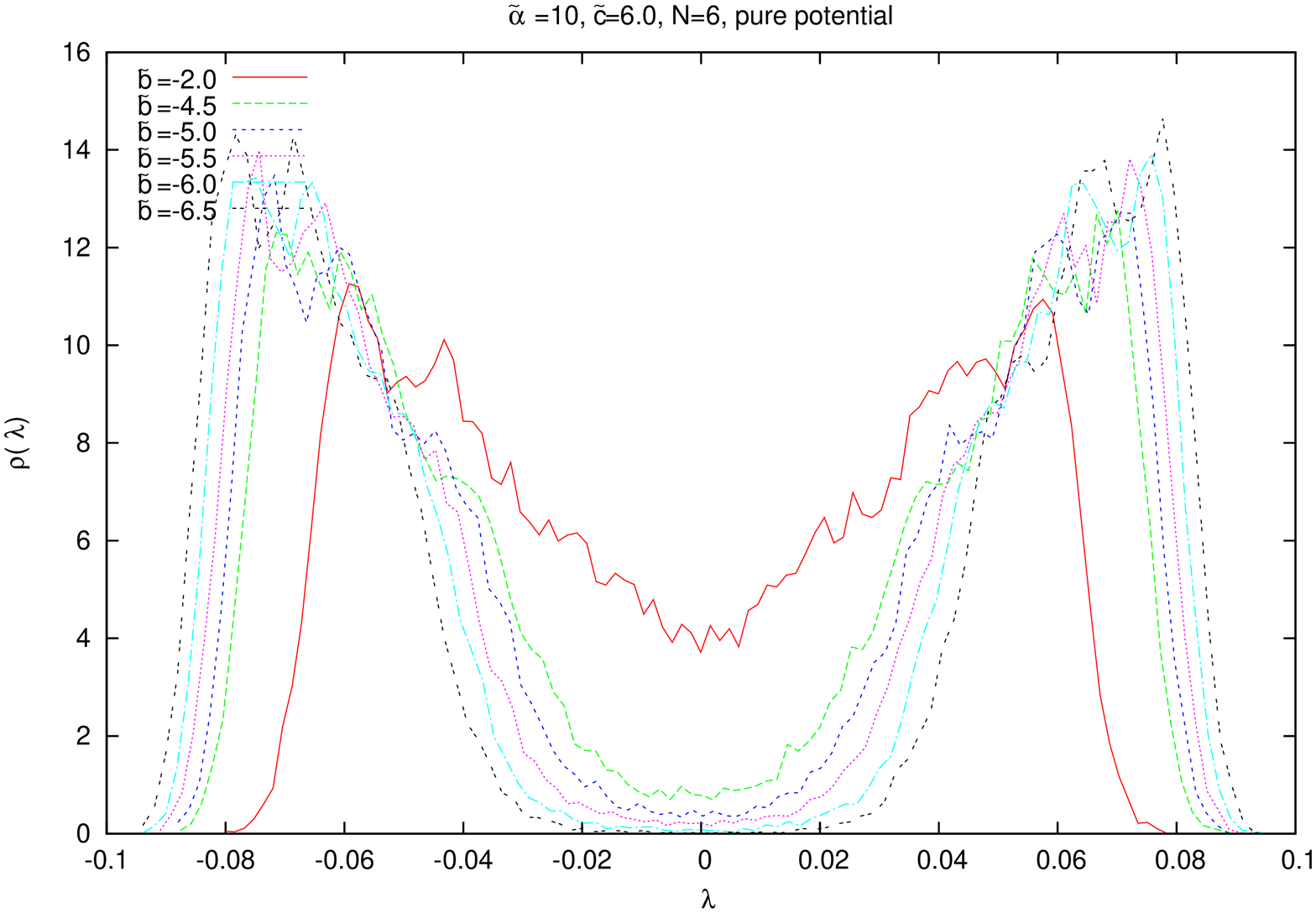}
\caption{The eigenvalue distributions across the non-uniform-to-disorder (matrix) transition in the full model and for the pure potential. }\label{ev_M}
\end{center}
\end{figure}

\paragraph{Eigenvalues Distributions and The Behavior Near the Triple Point:}
We have also investigated the matrix transition at the level of the eigenvalues distributions. In principle, the matrix transition occurs where the eigenvalues distributions split into two disjoint supports (cuts). In other words, it occurs at the point, where the distribution goes from a symmetric one centered around $0$ (as opposed to being centered around either $+\sqrt{-r/2u}$ or  $-\sqrt{-r/2u}$ in the case of the non-uniform phase), to a distribution with two symmetric cuts centered respectively around $\sqrt{-r/2u}$ and $-\sqrt{-r/2u}$. A sample of the eigenvalues distributions, in the full model and in the pure potential model, are shown on figure (\ref{ev_M}) for $N=6$ and $\tilde{c}=6$. 

We have used the eigenvalues distributions of $\Phi$, as the primary set of order parameters, employed in the determination of the matrix transition point, for smaller values of the coupling constant $\tilde{c}$. Following \cite{GarciaFlores:2009hf}, we have considered the regime $[2,3]$. This is the regime of interest to the calculation of the triple point (more on this below). We note that, the method employed above (maximum of magnetization and zero power), becomes unpractical in this regime. The results obtained for $N=4,6,10$ are included in table   (\ref{table1e}), and compared to the estimation of \cite{GarciaFlores:2009hf}.

We have determined the matrix transition point according to the following (somewhat arbitrary) criterion. We  have looked for the value of $\tilde{b}$, for which the eigenvalues distribution $\rho(\lambda)$ at $\lambda=0$, drops below $1$. The transition point is taken as the arithmetic average of this value of $\tilde{b}$, and the next one, for which, typically, the eigenvalues distribution at $\lambda=0$ becomes distinctly below $1$. 
A similar technique, to determine the matrix transition point, is employed in the recent thesis \cite{Vachovski}. 

\begin{figure}[htbp]
\begin{center}
\includegraphics[width=5.9cm,angle=-90]{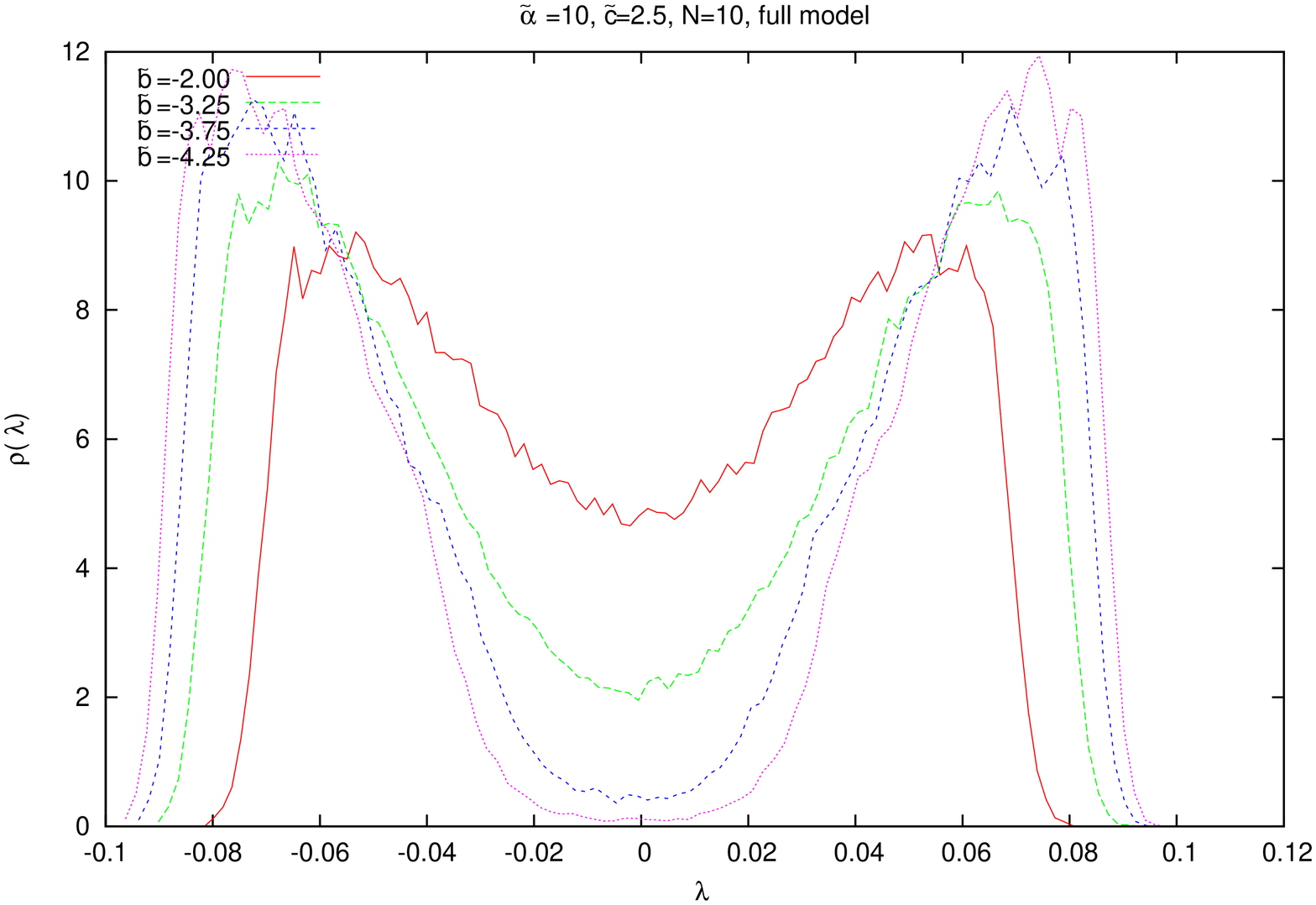}
\includegraphics[width=5.9cm,angle=-90]{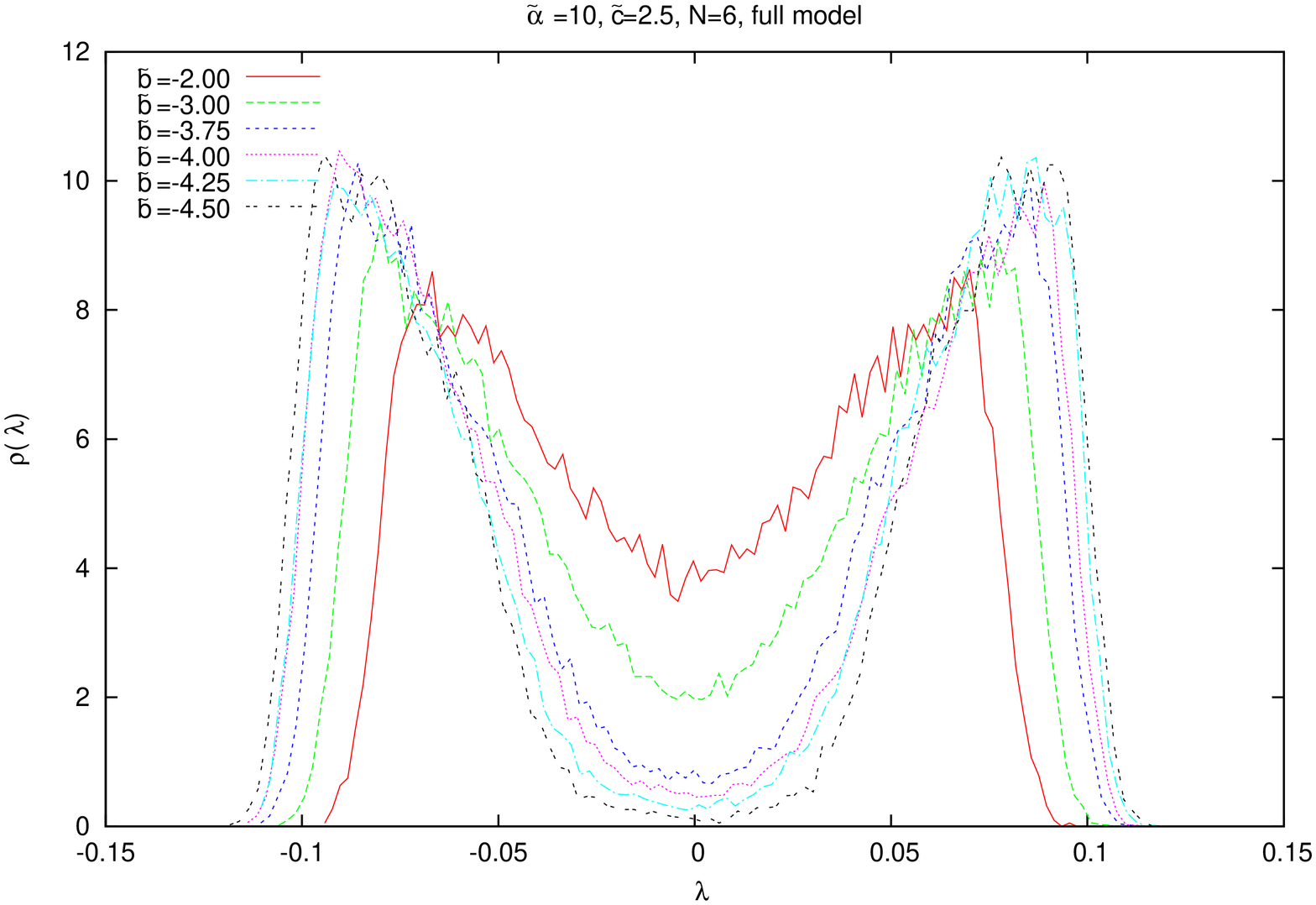}
\includegraphics[width=5.9cm,angle=-90]{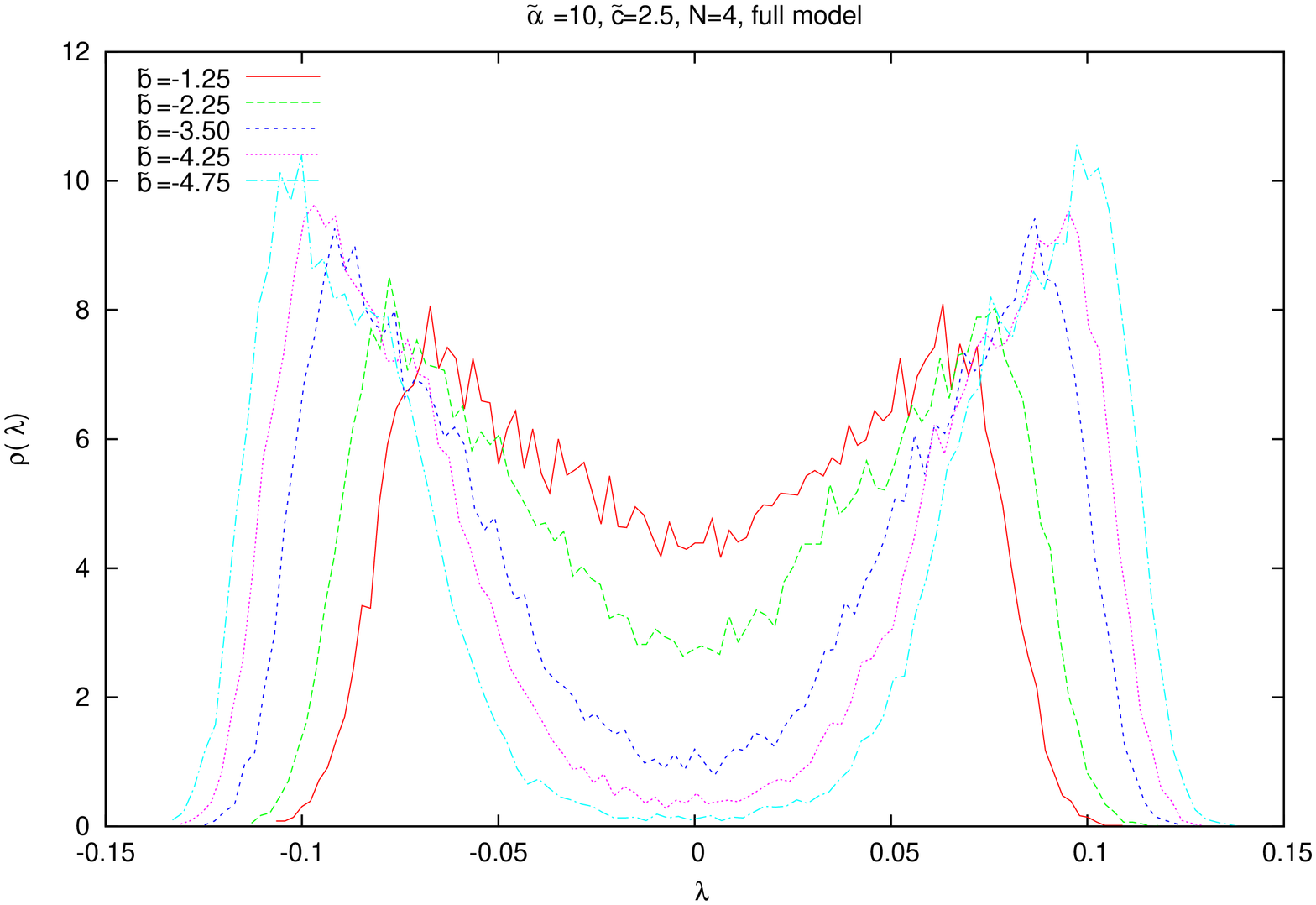}
\includegraphics[width=5.9cm,angle=-90]{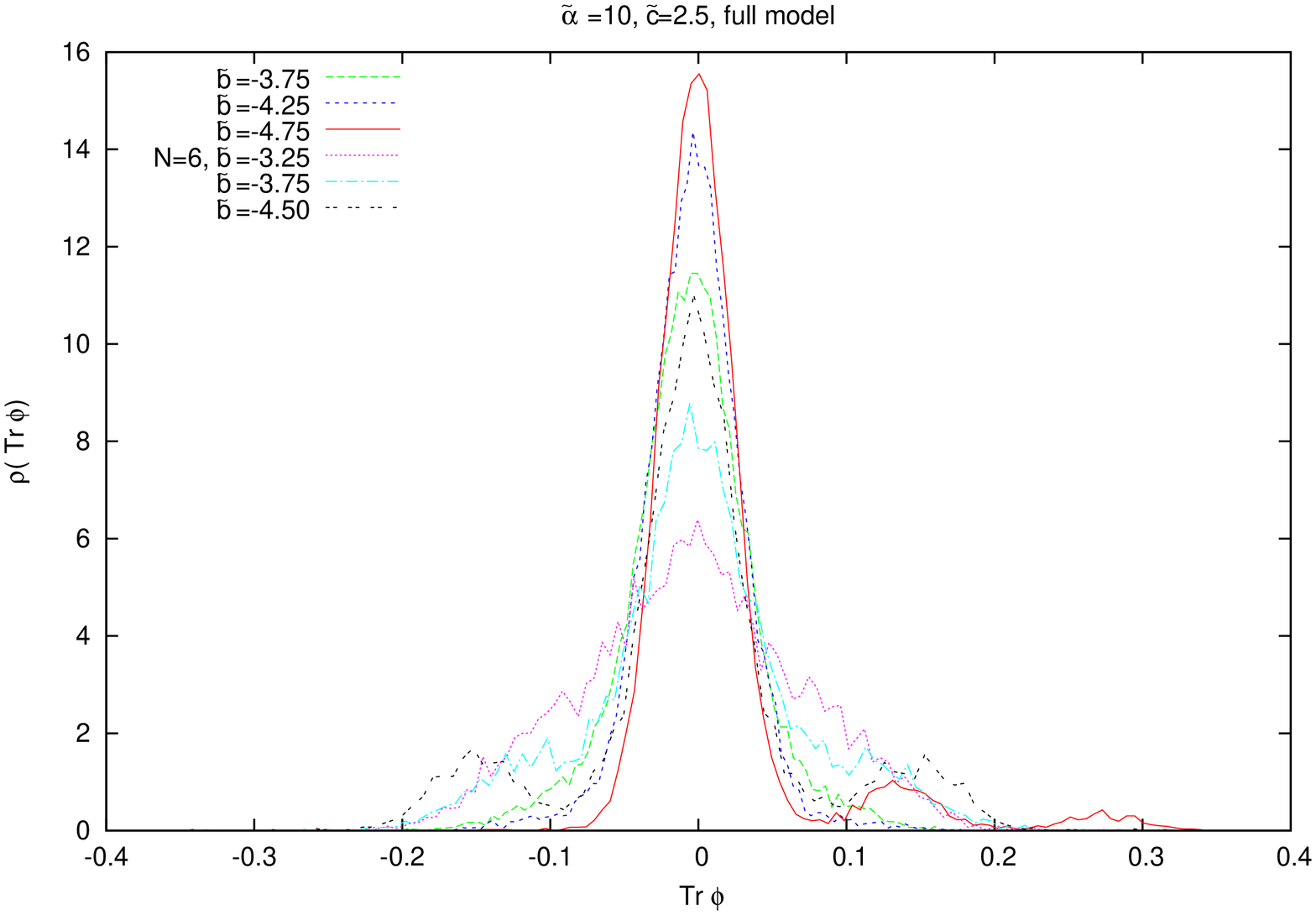}
\caption{The eigenvalues distributions across the non-uniform-to-disorder (matrix) transition in the full model. 
}\label{ev_Me}
\end{center}
\end{figure}

A sample of the eigenvalues distributions of $\Phi$ is shown on figure (\ref{ev_Me}) for   $\tilde{c}=2.5$. We also include, a sample of the probability distribution of ${\rm Tr}\Phi$, which may be used to determine the actual content of a given configuration $\Phi$. The number of pluses and minuses, can only be inferred, from the plot of the  probability distribution of ${\rm Tr}\Phi$. If $\Phi$ is a fluctuation about $0$ or $\gamma$, then the  probability distribution of ${\rm Tr}\Phi$ will contain a single symmetric peak around $0$.   There is also the possibility that  $\Phi$ is a fluctuation about $\gamma_k$, then the  probability distribution of ${\rm Tr}\Phi$ will contain a single symmetric peak around  $\sqrt{-r/2u}(2k-N)$. Typically, $\Phi$ will fluctuate  about  a mixed state, and as a consequence, several peaks will be present in the  probability distribution of ${\rm Tr}\Phi$ . For example, if $\Phi$ is a  mixture of $\gamma_k$ and $\gamma$, then, two peaks centered around $\sqrt{-r/2u}(2k-N)$ and $0$ will be present. Some examples are shown on figure (\ref{ev_Me}).

Using the results shown in table  (\ref{table1e}),  we can determine the non-uniform-ordered-to-disordered boundary. The fit to the  matrix (non-uniform-ordered-to-disordered) transition line is given by
 \begin{eqnarray}
\tilde{c}=(-1.3\pm 0.22)\tilde{b}-2.66\pm 0.9.\label{matrix}
\end{eqnarray}
This line is slightly different from the one measured in \cite{GarciaFlores:2009hf}, which may be due to our criterion for determining the matrix transition point. However, we should also recall that their result was obtained using a modification of the Metropolis algorithm which breaks detailed balance.

\begin{table}[h]
\centering
\begin{tabular}{|l|c|c|c|c|c|c| }
\hline
$\tilde{c}$ &  $N=4$ &  $N=6$ & $N=10$ &  $\tilde{b}_* ({\rm arithmetic}~{\rm average})$ & $\tilde{b}({\rm FDX})$   \\
\hline 
$3.0$ &  $-4.38 \pm 0.13$ & $-4.38\pm 0.13$ & $-4.38 \pm 0.13$ & $-4.38 \pm 0.13$ & $-3.38$ \\
$2.5$ &  $-3.88\pm 0.38$ & $-4\pm 0.25 $ & $-3.5\pm 0.25$ & $-3.79\pm 0.29$ & $-3.16$ \\
$2.0$ &  $-3.63 \pm 0.13 $ & $-3.63 \pm 0.13$ & $-3.63 \pm 0.13$ & $-3.63\pm 0.13 $ & $-2.94$ \\
\hline 
\end{tabular}
\caption{The matrix transition points near the triple point. }\label{table1e}
\end{table}


\subsection{Triple Point and Phase Diagram}  

The  most reliable estimation of the triple point can be obtained from the intersection point of (\ref{ising}) and (\ref{matrix}), because these two lines are the easiest, and the most accurate, to obtain with our gauge fixed Metropolis algorithm, and also with the algorithm of  \cite{GarciaFlores:2009hf}. In fact, they can even be accessed using the plain Metropolis algorithm. We deduce immediately that the triple point is located at
 \begin{eqnarray}
{\rm Ising}-{\rm matrix}~{\rm intersection}~{\rm point}~:~\tilde{b}_T
=-2.42~,~
\tilde{c}_T
=0.48.
\end{eqnarray}
Another estimation can be obtained from the intersection point of the matrix and the non-uniform lines. We get
 \begin{eqnarray}
{\rm non}-{\rm uniform}-{\rm matrix}~{\rm intersection}~{\rm point}~:~\tilde{b}_T
=-2.56~,~
\tilde{c}_T
=0.67.
\end{eqnarray}
These should be compared with the value $(-2.3,0.52)$ found in \cite{GarciaFlores:2009hf}. A natural candidate for the actual value of the triple point is, thus, the average value of the above two estimates, viz
\begin{eqnarray}
{\rm triple}~{\rm point}~:~\tilde{b}_T
=-2.49~,~
\tilde{c}_T
=0.58.
\end{eqnarray}
The error bars can be given by the rectangle with center given by the triple point $(-2.49,0.58)$, and corners given by the two intersection points $(-2.42,0.48)$, $(-2.56,0.67)$, and the two points $(-2.56,0.48)$, $(-2.42,0.67)$.

\begin{figure}[htbp]
\begin{center}
\includegraphics[width=13.0cm,angle=-90]{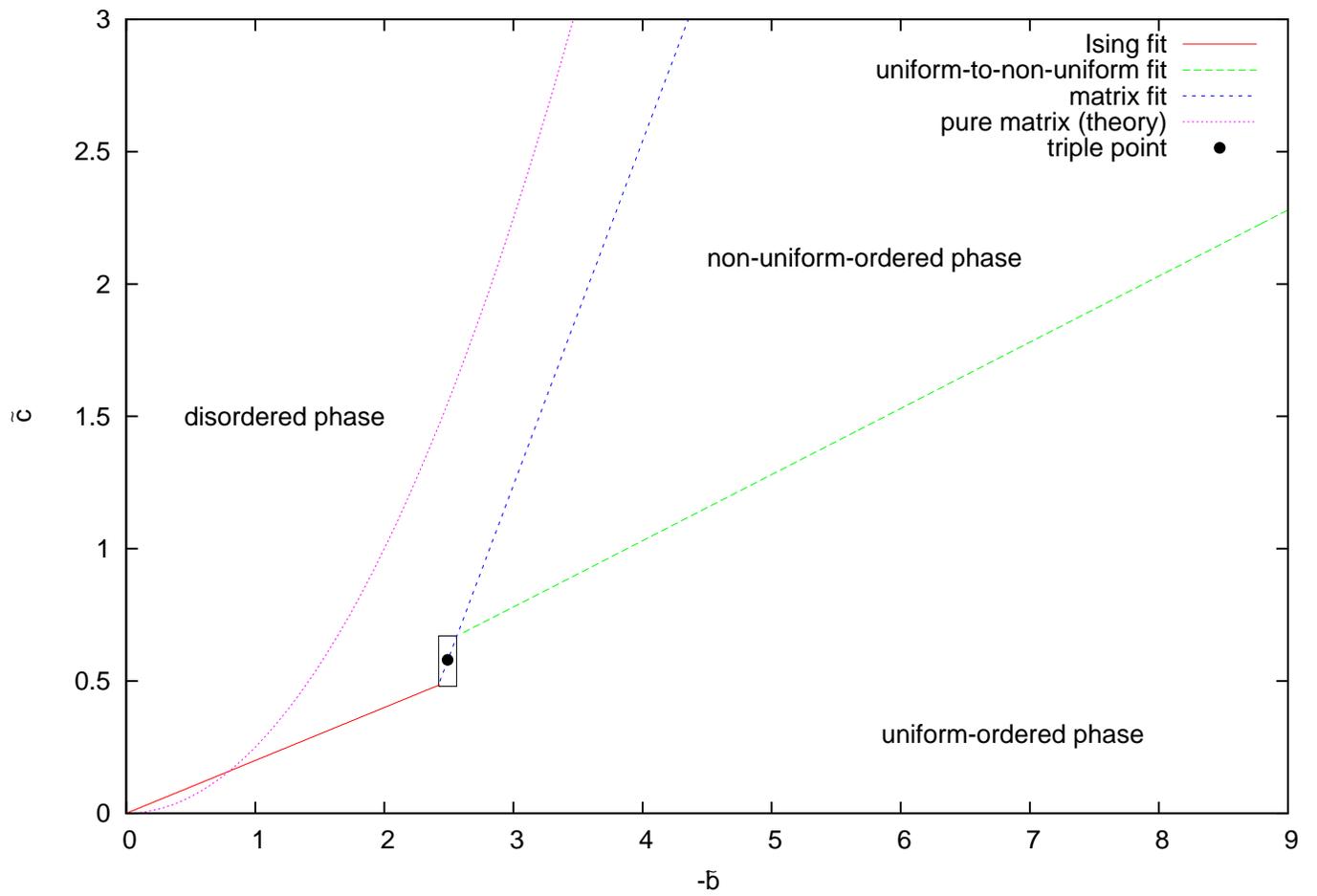}
\caption{The phase diagram. }\label{phase_diagram}
\end{center}
\end{figure}

\begin{figure}[htbp]
\begin{center}
\includegraphics[width=13.0cm,angle=-90]{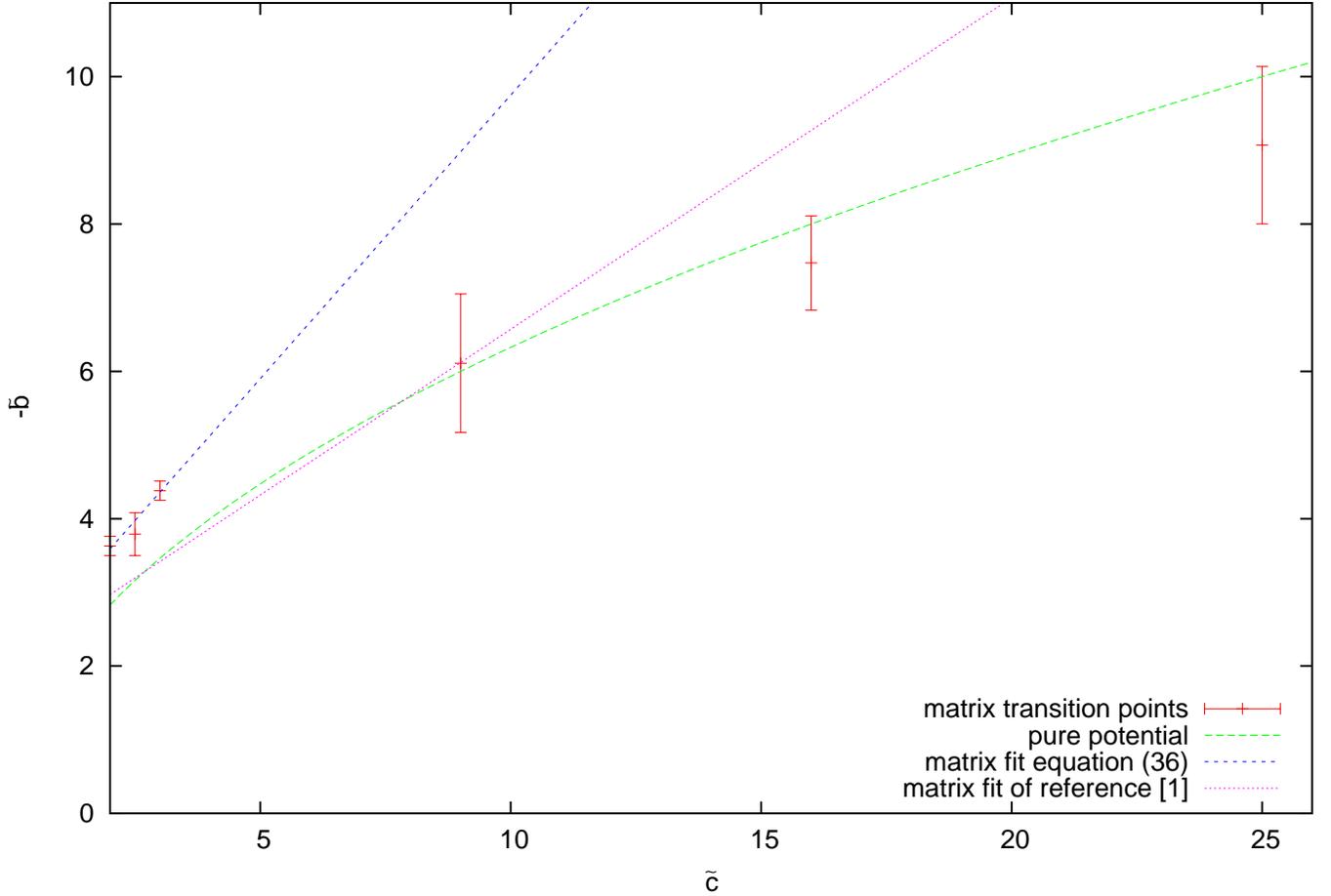}
\caption{The matrix (disorder-to-non-uniform-order) transition points for large and small values of the quartic coupling constant $\tilde{c}$. For large values of $\tilde{c}$, the Monte Carlo measurements converge to the prediction of the pure potential model. For small values of $\tilde{c}$, the fit is the straight line given by equation (\ref{matrix}), which must be extrapolated to even smaller values of $\tilde{c}$, in order to deduce an estimation of the triple point. We also compare, for small values  of $\tilde{c}$,  with the measurement of \cite{GarciaFlores:2009hf}. The discrepancies between the two measurements, for small $\tilde{c}$, is stemming from our criterion, based on the eigenvalues distributions, for determining the location of the matrix transition, which is different from the one used in  \cite{GarciaFlores:2009hf}.  }\label{phase_diagram1}
\end{center}
\end{figure}

\begin{figure}[htbp]
\begin{center}
\includegraphics[width=13.0cm,angle=-90]{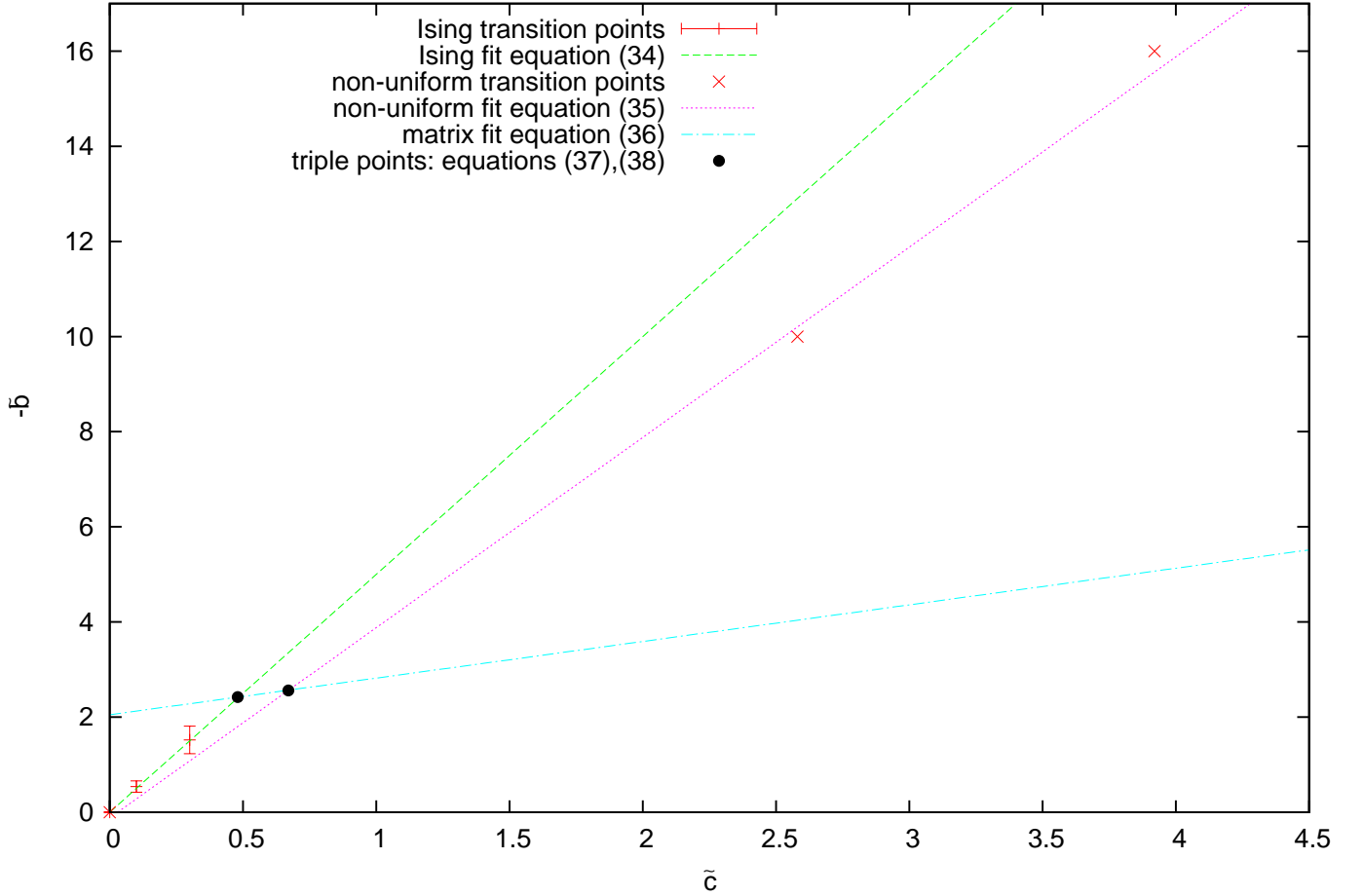}
\caption{The Ising (disorder-to-uniform) and the non-uniform-to-uniform transition lines. The Ising transition appears for  small values of $\tilde{c}$, while the non-uniform-to-uniform transition appears for large values of $\tilde{c}$. The Monte Carlo measurements of the Ising transition is fully consistent: more data points can be included quite easily, error bars are under control, large $N$ extrapolation is straightforward, and result obtained by our algorithm coincides with  the measurement of \cite{GarciaFlores:2009hf}. On the other hand, the two Monte Carlo measurements of the  non-uniform-to-uniform transition, included in this graph,   required much more calculation than their Ising and matrix counterparts put together. We did not attempt, here, to determine their error bars. The measured slope and small intercept, of the resulting non-uniform-to-uniform fit, are reasonably close to the measurements of \cite{GarciaFlores:2009hf}. Work on this major problem, i.e. a fully consistent determination of the non-uniform-to-uniform transition line,  is still in progress. We also plot the matrix line where the intersection points, with the Ising and the non-uniform-to-uniform lines, provide our two estimations of the triple point. }\label{phase_diagram0}
\end{center}
\end{figure}

The phase diagram is shown on figure (\ref{phase_diagram}). See also figures (\ref{phase_diagram1}) and (\ref{phase_diagram0}), where a close-up look at the matrix, the Ising, and the non-uniform transition lines is shown.

\subsection{Comparison of Various Algorithms}
The algorithm used in \cite{GarciaFlores:2009hf} to compute the phase diagram is based on, a very complex variation, of the Metropolis algorithm, which does not preserve detailed balance. In the region of the disordered phase, their algorithm behaves essentially as the usual Metropolis algorithm, with a processing time per configuration, with respect to the matrix size, proportional to $N^4$. The new Metropolis algorithm, described in   \cite{GarciaFlores:2009hf}, behaves better and better, as we go farther and farther, from the origin, i.e. towards the regions of the uniform and non-uniform phases. The processing time per configuration, with respect to the matrix size, is claimed to be proportional to $N^3$, for the values of $N$ between $4$ and $64$. See graph $9.12$ of F.G Flores' doctoral thesis\footnote{Not available on the ArXiv.}, where we can fit this region of $N$ with a straight line.  Also, it is worth noting, that this new algorithm involves, besides the usual optimizable parameters found in the Metropolis algorithm, such as the acceptance rate, a new optimizable parameter $p$, which controls the compromise between the speed and the accuracy of the algorithm. For $p=0$ we have a fast process with considerable relative systematic error, while for $p=1$ we have a slow process but a very small relative error. This error is, precisely, due to the lack of detailed balance. Typically we fix this parameter around $p=0.55-0.7$.

The algorithm of \cite{GarciaFlores:2009hf} is the only known method, until now, which is successful in mapping the complete phase diagram of noncommutative phi-four on the fuzzy sphere. However we had found it, from our experience, very hard to reproduce this work. 

Our first original goal was to find an alternative method which is, $i)$ conceptually as simple as the usual Metropolis method, and $ii)$ without systematic errors, and $iii)$ can map the whole phase diagram. This goal was achieved by the algorithm described and used in this article.  The processing time per configuration, with respect to the matrix size, in our algorithm, is proportional to $N^4$, which is comparable to the usual Metropolis algorithm, but with the virtue that we can access the non-uniform phase. There is no systematic errors in this algorithm, and hence no analogue of the parameter $p$ mention above. 

How does our algorithm compares with the algorithm of \cite{GarciaFlores:2009hf}, is a much harder question, since we have no complete understanding of the detail of their algorithm. Their algorithm is faster, but this can not be the only concern. Accuracy of the method, and conceptual simplicity, are also very important virtues, especially, for difficult problems, such as this one, where the physics is extremely interesting, but very hard to attain. Our algorithm satisfies both these two requirements.

Our other goal, in this article, was to compare the results obtained by the two methods for the non-uniform phase. There are still discrepancies between the two methods which is very puzzling. The non-uniform phase is characterized, in this article,  by the "discontinuity/jump" in the expectation value of the kinetic term, the total power, power in the zero mode and magnetization. According to  \cite{GarciaFlores:2009hf}, this jump is also associated with a peak in the susceptibility and specific heat indicative of a second-order behavior, which is something we were not able to reproduce in our scheme, in any consistent way. This is very troubling, to say the least, because we could not, from what we have and know at this point, ascertain whether this is due to a technical problem, or if it is a genuine discrepancy.

\section{The Self-Dual Noncommutative $\Phi^4$ on the Fuzzy Sphere}

\subsection{Self-Dual Noncommutative $\Phi^4$}
We consider, for simplicity, a real scalar field on the noncommutative (Moyal-Weyl) plane $[\hat{x}_\mu,\hat{x}_\nu]=i{\theta}_{\mu\nu}$. The phi-four theory on the  noncommutative plane is, a particular limit, of a one-parameter family of phi-four models on the  noncommutative plane, obtained by the addition of an extra operator,  the harmonic oscillator potential , to the kinetic part of the action.  The action reads explicitly
\begin{eqnarray}
S_{\Omega}
&=&\sqrt{{\rm det} \pi{\theta}}~{\rm Tr}_H\bigg[-\frac{1}{2}\hat{\phi}~\hat{\partial}_{\mu}^2\hat{\phi}+\frac{\Omega^2}{2\theta^2}\{\hat{x}_{\mu},\hat{\phi}\}^2+\frac{\mu^2}{2}\hat{\phi}^2+\frac{\lambda}{4!}\hat{\phi}^4\bigg].
\end{eqnarray}
We know that derivations on ${\bf R}^2_{\theta}$ are inner, given by the adjoint action, viz 
\begin{eqnarray}
\hat{\partial}_{\mu}\hat{\phi}=\frac{1}{i}({\theta}^{-1})_{\mu\nu}[\hat{x}_{\nu},\hat{\phi}].
\end{eqnarray}
Alternatively, the action can be rewritten as
\begin{eqnarray}
S_{\Omega}
&=&\sqrt{{\rm det} \pi{\theta}}~{\rm Tr}_H\bigg[\frac{1+\Omega^2}{\theta^2}\hat{x}_{\mu}^2\hat{\phi}^2-\frac{1-\Omega^2}{\theta^2}\hat{x}_{\mu}\hat{\phi}\hat{x}_{\mu}\hat{\phi}+\frac{\mu^2}{2}\hat{\phi}^2+\frac{\lambda}{4!}\hat{\phi}^4\bigg].
\end{eqnarray}
This is the  Grosse-Wulkenhaar model. The addition of the harmonic oscillator potential to the kinetic action modifies, and thus allows us to control, the IR behavior of the theory. A particular version of this theory was shown to be renormalizable by Grosse and Wulkenhaar in \cite{Grosse:2004yu}. It was shown in  \cite{Langmann:2002cc}, that this action is covariant under a duality transformation which exchanges, among other things, positions and momenta. The value ${\Omega}^2=1$, in particular, gives an action which is invariant under this duality transformation. The theory at  ${\Omega}^2=1$ is called the Langmann-Szabo model or the  self-dual Grosse-Wulkenhaar model.

The usual  phi-four theory on the  noncommutative plane corresponds to the limit $\Omega\longrightarrow 0$. The other interesting limit is $\Omega\longrightarrow 1$, which corresponds to the  self-dual Grosse-Wulkenhaar model. The main technical simplification, occurring in the limit  $\Omega\longrightarrow 1$, is the observation that the off-diagonal term in the action drops, and we end up with the action\footnote{After regularization this action becomes the Penner matrix model.} 

\begin{eqnarray}
S_{\Omega=1}
&=&\sqrt{{\rm det} \pi{\theta}}~{\rm Tr}_H\bigg[\frac{2}{\theta^2}\hat{x}_{\mu}^2\hat{\phi}^2+\frac{\mu^2}{2}\hat{\phi}^2+\frac{\lambda}{4!}\hat{\phi}^4\bigg].
\end{eqnarray}
Let us now introduce creation and annihilation operators $a^+$ and $a$ satisfying $[a,a^+]=\theta$ by
\begin{eqnarray}
\hat{x}_1=\frac{1}{\sqrt{2}}(a+a^+)~,~\hat{x}_2=\frac{1}{i\sqrt{2}}(a-a^+).
\end{eqnarray}
The number operator $\hat{N}$ is defined by $\hat{N}=a^+a/\theta$. We can verify, for example, that $\hat{x}_{\mu}^2=2\theta \hat{N}+\theta$. We will work in the number basis defined by
\begin{eqnarray}
\hat{N}|n>=n|n>~,~a^+|n>=\sqrt{\theta(n+1)}|n+1>~,~a|n>=\sqrt{\theta n}|n-1>.
\end{eqnarray}
The components of  $\hat{\phi}$, in the number basis, are given by $ \tilde{\phi}_{nm}=<n-1|\hat{\phi}|m-1>$. In the number basis $\{|n>\}$ the action $S_{\Omega}$ reads explicitly  

\begin{eqnarray}
S_{\Omega}&=&r\sum_{m=1}^{\infty}\sum_{n=1}^{\infty}\tilde{\phi}_{mn}\tilde{\phi}_{nm}+u \sum_{m=1}^{\infty}\sum_{n=1}^{\infty}\sum_{k=1}^{\infty}\sum_{l=1}^{\infty}\tilde{\phi}_{mn}\tilde{\phi}_{nk}\tilde{\phi}_{kl}\tilde{\phi}_{lm}+\pi(1+\Omega^2)\sum_{m=1}^{\infty}\sum_{n=1}^{\infty}(m+n-1)\tilde{\phi}_{mn}\tilde{\phi}_{nm}\nonumber\\
&-&\pi(1-\Omega^2)~\sum_{m=1}^{\infty}\sum_{n=1}^{\infty}\bigg[\sqrt{(m-1)(n-1)}\tilde{\phi}_{mn}\tilde{\phi}_{n-1m-1}+\sqrt{mn}\tilde{\phi}_{mn}\tilde{\phi}_{n+1m+1}\bigg].
\end{eqnarray}
This is a three-parameter model, where the mass parameter $r$ and the quartic coupling $u$, are given by
\begin{eqnarray}
r=\pi\theta \frac{{\mu}^2}{2}~,~u=\pi\theta\frac{\lambda}{4!}.
\end{eqnarray}
The other coupling is the harmonic oscillator coupling $\Omega$.
\subsection{Fuzzy Sphere as a Regulator}
In the remainder of this section, we will write down a non-perturbative regularization of this theory on the fuzzy sphere. We only need to consider the kinetic term.  Let $L_a$ be the generators of $SU(2)$ in the irreducible representation of dimension $N$, i.e. $L_a$ are the angular momenta of spin $(N-1)/{2}$. In other words, $[L_a,L_b]=i{\epsilon}_{abc}L_c$, and $L_a^2=\frac{N^2-1}{4}=c_2$ is the quadratic Casimir. The noncommutativity parameter $\theta$, on the fuzzy sphere, is defined by $\theta=R^2/\sqrt{c_2}$, where  $R$ is the radius of the sphere\footnote{By sitting on the north pole, i.e. $\hat{x}_3=R {\bf 1}_N$, and taking the limit $N\longrightarrow\infty$, and $R\longrightarrow\infty$, keeping $R^2/\sqrt{c_2}=\theta$ fixed, the fuzzy sphere reduces to the noncommutative plane.}. The derivatives, and the round Laplacian on the fuzzy sphere are defined by 
\begin{eqnarray}
{\cal L}_a=\frac{i}{R}[L_a,...].
\end{eqnarray}
\begin{eqnarray}
\Delta_0={\cal L}_a^2.
\end{eqnarray}
We will work in the basis $\{|m>\}$ defined by the usual relations $L_3|m>=m|m>$, $L_{\pm}|m>=\sqrt{l(l+1)-m(m\pm 1)}|m\pm 1>$, where $l=(N-1)/2$ and $L_{\pm}=L_1\pm iL_2$. We relabel the basis as $|m>=|i>$, where $m=i-l-1$. We compute $(L_3)_{ij}={\delta}_{ij}(2i-N-1)/2$,  $(L_+)_{ij}=\sqrt{j(N-j)}\delta_{i-1,j}$, $(L_-)_{ij}=\sqrt{i(N-i)}\delta_{i+1,j}$. Rotating around the $x$-axis, with an angle $\pi$, we have $L_1\longrightarrow L_1^{'}=L_1$, $L_2\longrightarrow L_2^{'}=-L_2$, i.e. $L_{\pm}\longrightarrow  L_{\pm}^{'}=L_{\mp}$, and $L_3\longrightarrow L_3^{'}=-L_3$. Thus $(L_3^{'})_{ij}={\delta}_{ij}(N+1-2i)/2$,  $(L_-^{'})_{ij}=\sqrt{j(N-j)}\delta_{i-1,j}$, $(L_+^{'})_{ij}=\sqrt{i(N-i)}\delta_{i+1,j}$.

A real scalar field $\hat{\phi}$ is a hermitian $N\times N$ matrix which will be expanded in the obvious way
\begin{eqnarray}
\hat{\phi}=\sum_{m_1=-l}^{l}\sum_{m_2=-l}^l\hat{\phi}_{m_1m_2}|m_1><m_2|=\sum_{i=0}^{N-1}\sum_{j=0}^{N-1}\hat{\phi}_{ij}|i><j|~,~\hat{\phi}_{m_1m_2}\equiv \hat{\phi}_{ij}.
\end{eqnarray}
We start by considering  a more general Laplacian, obtained by adding a harmonic oscillator potential to $\Delta_0$, in the most obvious way. First, we introduce the coordinates operators $\hat{x}_a$, on the fuzzy sphere, by $\hat{x}_a={R L_a}/{\sqrt{c_2}}$, which satisfy $[\hat{x}_a,\hat{x}_b]=iR \epsilon_{abc}\hat{x}_c/\sqrt{c_2}$ and $\hat{x}_a^2=R^2$. We define the right-acting coordinate operators $\hat{x}_a^R$ by $\hat{x}_a^R\hat{\phi}=\hat{\phi}\hat{x}_a$, and then introduce the  coordinates operators $X_a$ by
\begin{eqnarray}
X_a=\frac{\hat{x}_a+\hat{x}_a^R}{2}.
\end{eqnarray}
We define the Laplacian
\begin{eqnarray}
\Delta_{\Omega}^{'}={\cal L}_a^2-\frac{4\Omega^2}{\theta^2} X_a^2.
\end{eqnarray}
In other words, we consider the kinetic term
\begin{eqnarray}
\frac{K}{4\pi R^2}&=&\frac{1}{2}{\rm Tr}_H\hat{\phi}(-\Delta_{\Omega}^{'})\hat\phi\nonumber\\
&=&\frac{1}{R^2}{\rm Tr}_H\bigg((1+\Omega^2)c_2\hat{\phi}^2-(1-\Omega^2)\hat{\phi}{L}_3\hat{\phi}{L}_3-(1-\Omega^2)\hat{\phi}{L}_+\hat{\phi}{L}_-\bigg).
\end{eqnarray}
The normalization $4\pi R^2$ is chosen such that in the commutative limit $N\longrightarrow\infty$ we have $(4\pi R^2){\rm Tr}_H/N\longrightarrow R^2\int_{S^2}d\Omega_2$. Explicitly, we compute (with $\hat{\phi}_{i-1j-1}=\tilde{\phi}_{ij}/\sqrt{2\pi}$)

\begin{eqnarray}
\frac{K}{4\pi R^2}&=&\frac{1+\Omega^2}{2\pi \theta}\sum_{i=1}^{N}\sum_{j=1}^{N}(i+j-1-\frac{2(i-1)(j-1)}{N-1})\tilde{\phi}_{ij}\tilde{\phi}_{ji}\nonumber\\
&-&\frac{1-\Omega^2}{2\pi \theta}\sum_{i=1}^{N}\sum_{j=1}^{N}\sqrt{(i-1)(j-1)(1-\frac{i-2}{N-1})(1-\frac{j-2}{N-1})}\tilde{\phi}_{ij}\tilde{\phi}_{j-1i-1}\nonumber\\
&-&\frac{1-\Omega^2}{2\pi \theta}\sum_{i=1}^{N}\sum_{j=1}^{N}\sqrt{ij(1-\frac{i-1}{N-1})(1-\frac{j-1}{N-1})}\tilde{\phi}_{ij}\tilde{\phi}_{j+1i+1}\nonumber\\
&-&\frac{\Omega^2}{2\pi \theta}\sum_{i=1}^{N}\sum_{j=1}^{N}(2i+2j-N-3-\frac{4(i-1)(j-1)}{N-1})\tilde{\phi}_{ij}\tilde{\phi}_{ji}.
\end{eqnarray}
The last  term is not present on the noncommutative plane which is, clearly, an unwanted effect. After some trial and error, we have discovered, that the correct Laplacian on the fuzzy sphere, which reproduces precisely the effect of the harmonic oscillator potential,  is given by
\begin{eqnarray}
\Delta_{\Omega}={\cal L}_a^2+\Omega^2{\cal L}_3^2-\frac{4\Omega^2}{\theta^2}(X_a^2-X_3^2).
\end{eqnarray}
This will describe a squashed fuzzy sphere, which  is more appropriate, for the non-perturbative description of the noncommutative plane. Indeed, we compute

\begin{eqnarray}
\frac{K}{4\pi R^2}&=&\frac{1}{2}{\rm Tr}_H\hat{\phi}(-\Delta_{\Omega})\hat\phi\nonumber\\
&=&\frac{1}{R^2}{\rm Tr}_H\bigg((1+\Omega^2)c_2\hat{\phi}^2-(1+\Omega^2)\hat{\phi}{L}_3\hat{\phi}{L}_3-(1-\Omega^2)\hat{\phi}{L}_+\hat{\phi}{L}_-\bigg).
\end{eqnarray}
Equivalently
\begin{eqnarray}
\frac{K}{4\pi R^2}&=&\frac{1+\Omega^2}{2\pi\theta}\sum_{i=1}^{N}\sum_{j=1}^{N}(i+j-1-\frac{2(i-1)(j-1)}{N-1})\tilde{\phi}_{ij}\tilde{\phi}_{ji}\nonumber\\
&-&\frac{1-\Omega^2}{2\pi \theta}\sum_{i=1}^{N}\sum_{j=1}^{N}\sqrt{(i-1)(j-1)(1-\frac{i-2}{N-1})(1-\frac{j-2}{N-1})}\tilde{\phi}_{ij}\tilde{\phi}_{j-1i-1}\nonumber\\
&-&\frac{1-\Omega^2}{2\pi \theta}\sum_{i=1}^{N}\sum_{j=1}^{N}\sqrt{ij(1-\frac{i-1}{N-1})(1-\frac{j-1}{N-1})}\tilde{\phi}_{ij}\tilde{\phi}_{j+1i+1}.
\end{eqnarray}
We can now include a mass term and a phi-four coupling in a trivial way. The full action, on the fuzzy sphere, will read
\begin{eqnarray}
S_{\Omega}&=&4\pi R^2{\rm Tr}_H\bigg[-\frac{1}{2}\hat{\phi}\bigg({\cal L}_a^2+\Omega^2{\cal L}_3^2-\frac{4\Omega^2}{\theta^2}(X_a^2-X_3^2)\bigg)\hat{\phi}+\frac{\mu^2}{2}\hat{\phi}^2+\frac{\lambda}{4!}\hat{\phi}^4\bigg].
\end{eqnarray}
We scale the field as $\hat{\phi}=\tilde{\phi}/\sqrt{2\pi}$, and also introduce  the parameters
\begin{eqnarray}
r=\mu^2 R^2~,~u=\frac{\lambda R^2}{4!\pi}.
\end{eqnarray}
The full action can then be rewritten as

\begin{eqnarray}
S_{\Omega}&=&{\rm Tr}_H\bigg[- [L_a,\tilde{\phi}]^2- \Omega^2[L_3,\tilde{\phi}]^2+ \Omega^2\{L_a,\tilde{\phi}\}^2 +r\tilde{\phi}^2+u\tilde{\phi}^4\bigg].
\end{eqnarray}
This is a one-parameter family of phi-four models on the fuzzy sphere which generalizes (\ref{fundamental}). Coupling to a $U(1)$ gauge field is straightforward, i.e. we make the replacement $L_a\longrightarrow \sqrt{N/2}X_a$. The analogue of (\ref{Sm}) is obviously given by
\begin{eqnarray}
S_{\Omega}=-\frac{N}{2}Tr[X_a,\tilde{\phi}]^2-\frac{N\Omega^2}{2}Tr[X_3,\tilde{\phi}]^2+\frac{N\Omega^2}{2}Tr\{X_a,\tilde{\phi}\}^2+Tr V(\tilde{\phi}).\label{Sm1}
\end{eqnarray}

\section{Conclusion and Outlook}

In this article, we have proposed a new algorithm for the Monte Carlo simulation of noncommutative phi-four on the fuzzy sphere, and also reported our first numerical results on the corresponding phase diagram, obtained with small values of $N$ up to $N=10$, and large numbers of statistics. Basically, the new algorithm employs gauge invariance  in order to reduce the scalar sector to the core eigenvalues problem. The phase diagram is complex consisting of three transition lines: the Ising or uniform-to-disorder, the matrix or non-uniform-to-disorder, and the uniform-to-non-uniform transition lines. These lines  intersect at a triple point. The measurement of the uniform-to-non-uniform transition line, using our algorithm, remains very demanding but tractable. The measurements, included in this article, are largely consistent with those reported originally in \cite{GarciaFlores:2009hf}.

The first immediate extension of this work is to optimize the algorithm further, and push the calculation of the phase diagram to higher values of $N$, with reasonably large numbers of statistics, especially in the case of  the uniform-to-non-uniform transition line. We note that a major improvement of our algorithm,  may be achievable,  by replacing the Metropolis updating procedure, for the scalar eigenvalues problem, by the Hybrid Monte Carlo algorithm, whereas we may keep using the very efficient Metropolis for the gauge sector. 

Another immediate line of investigation is the calculation of the phase diagram of the self-dual  noncommutative phi-four on the fuzzy sphere, constructed in the last section. The main question, here, is what happens to the Ising transition line, as $\Omega$ goes from $\Omega=0$ to $\Omega=1$, and as a consequence,  what is the fate of the triple point.

\paragraph{Acknowledgments:} This research was supported by "The National Agency for the
Development of University Research (ANDRU), MESRS, Algeria", under PNR contract number U23/Av58
(8/u23/2723). The Monte Carlo simulations, reported in this article, were largely performed  on the machines of the School of Theoretical Physics, Dublin Institute for Advanced Studies, Dublin, Ireland.

\newpage



\end{document}